\documentclass{aa}  

\usepackage{graphicx}
\usepackage{txfonts}
\usepackage{lipsum}
\usepackage[normalem]{ulem}
\usepackage[colorlinks=true, linkcolor=blue, citecolor=blue, urlcolor=blue]{hyperref}
\usepackage{amsmath}
\usepackage{amssymb}
\usepackage{mathtools}
\usepackage{xcolor}

\begin{document}

   \title{Linking high-mass X-ray binaries to binary compact object mergers in Milky Way-like galaxies}
   \titlerunning{Linking high-mass X-Ray binaries to binary compact object mergers}
   
   \author{Felipe Vivanco C\'adiz
          \inst{1}
          \and
          M. Celeste Artale
          \inst{1}
          \and 
          Nicola Masetti\inst{2,1}
          \and
          Gaston Escobar\inst{3,4}
          \and
          Giuliano Iorio\inst{5}
          \and
          Ankit Kumar\inst{1}
          \and
          Boyuan Liu\inst{6}
          \and
          Michela Mapelli\inst{6,7,8,9}
    }
   \authorrunning{Vivanco Cádiz et al.}
   \institute{Universidad Andres Bello, Facultad de Ciencias Exactas, Departamento de F\'isica y Astronom\'ia, Instituto de Astrof\'isica, Fernandez Concha 700, Las Condes, Santiago RM, Chile\\
    \email{fvivanco@me.com}, \email{maria.artale@unab.cl}
    \and
    INAF - Osservatorio di Astrofisica e Scienza dello Spazio, via Piero Gobetti 101, I-40129 Bologna, Italy
    \and
    Instituto de Astrofísica de Canarias, E-38205 La Laguna, Tenerife, Spain
    \and
    Universidad de La Laguna, Dpto. Astrofísica, E-38206 La Laguna, Tenerife, Spain
    \and
    Institut de Ciències del Cosmos (ICCUB), Universitat de Barcelona (UB), c. Martí i Franquès 1, 08028 Barcelona, Spain
    \and
    Universität Heidelberg, Zentrum für Astronomie (ZAH), Institut für Theoretische Astrophysik, Albert Ueberle Str. 2,
    69120 Heidelberg, Germany
    \and
    Universität Heidelberg, Interdisziplinäres Zentrum für Wissenschaftliches Rechnen, 69120 Heidelberg, Germany
    \and
    Dipartimento di Fisica e Astronomia Galileo Galilei, Università di Padova, Vicolo dell’Osservatorio 3, 35122 Padova, Italy
    \and
    INFN, Sezione di Padova, Via Marzolo 8, 35131 Padova, Italy}

   \date{Received ...; accepted ...}

    \abstract{High-mass X-ray binaries (HMXBs) constitute a potential intermediate phase connecting massive stellar binaries to the formation of coalescing binary compact objects (BCOs). Yet, the specific circumstances that allow HMXBs to later develop into merging BCOs are still under debate.
    In this study, we focus on \textit{wind-fed} HMXBs and investigate this link by generating synthetic catalogs with the population-synthesis code {\sc{sevn}} and assigning them to a sample of 66 Milky Way-like galaxies. Using stellar particles drawn from the TNG50 cosmological simulation, we assign HMXBs according to particle mass, age, and metallicity, and follow their subsequent evolution to BCOs.
    In this way, our approach provides a realistic framework for modelling HMXBs in the Milky Way by explicitly accounting for the varied metallicities and spatial distributions of binary systems.
    Our method recovers the spatial distribution inferred from the observed properties of Galactic HMXBs: their positions follow the spiral arms and agree with the measured radial profile. The age distribution aligns well with observations and shows that BH-HMXBs tend to be younger than NS-HMXBs.
    We find that the fraction of HMXBs that evolve into merging BCOs within a Hubble time for BH-HMXBs is $\sim0.2$–$3.2\%$ (or $\sim0.3$–$5.4\%$ when restricting to $10^{35}\leq L_X\leq10^{40}\,$erg\,s$^{-1}$), while for NS–HMXBs is $\sim3.6$–$23.4\%$ (or $\sim3.5$–$26.0\%$ with the same luminosity cut). We show that common envelope (CE) episodes and natal-kick magnitudes are the primary processes determining the number of merging systems. Successful BCO mergers are typically produced by systems that experienced significant orbital hardening (early CE), whereas stable mass transfer often yields wide, non-merging binaries unless extreme eccentricities are induced by kicks. Metallicity plays a secondary but important role by modulating compact-remnant masses and wind-driven orbital widening.}

   \maketitle
%

\section{Introduction}

Massive stars often form in binary or multiple systems (see e.g., \citealt{Offner2023}), where mass transfer and stellar winds shape their subsequent evolution and final fate \citep{Sana2012}. These interactions give rise to a variety of systems. Among these, high-mass X-ray binaries (HMXBs) constitute a potential intermediate phase connecting massive stellar binaries to the formation of binary compact objects (BCOs). 

HMXBs are binary systems composed of a compact object (either a black hole, BH, or a neutron star, NS) and a high-mass OB-type star with masses larger than \(8\ \mathrm{M_{\odot}}\).
Depending on the binary properties, this mass transfer occurs via stellar wind capture or Roche-lobe overflow (RLOF, \citealt{Eggleton1983}). Several works show that HMXBs are highly sensitive to metallicity, with lower metallicities producing more numerous systems and brighter X-ray populations \citep{Dray2006,Mapelli2009, Michela2010,Linden2010,Prestwich2013,Antoniou2016,Lehmer2022,Lehmer2024}. On the other hand, their relatively short lifetimes, ranging from \(\sim2\ \mathrm{to}\  35\)~Myr after the formation of the binary, link HMXBs closely to recent star formation  \citep[see e.g.,][]{Grimm2003}.
Together, these properties suggest that the HMXB population in a galaxy is significantly influenced by its star formation activity and stellar chemical composition. Observational studies have established a correlation between the total X-ray luminosity of HMXBs and the host galaxy star formation rate \citep[see e.g.,][]{Mineo2012}, while metal-poor galaxies tend to host a larger number of HMXBs than their metal-rich counterparts (see e.g., \citealt{Douna2015}).

A fraction of the HMXB population is expected to evolve into BCOs, as the massive companion star may collapse into a second compact object while remaining gravitationally bound to the first. This process can produce binary black holes (BBHs), binary neutron stars (BNSs), or black hole–neutron star binaries (BHNSs). If the BCOs eventually merge, these objects would produce gravitational wave (GW) events which could be observed by GW detectors \citep{Abbot2017,Abbot2019,Abbot2021a,Abbot2021b,ET2025_bluebook}. However, whether HMXBs significantly contribute to the merging BCO population remains uncertain. Several physical processes can disrupt the binary evolution pathway: the donor star and compact object may merge earlier, the system can be unbound during the supernova event from the companion, or the resulting BCO may retain a separation too wide to merge within a Hubble time. Understanding which evolutionary outcomes dominate is therefore essential for assessing the contribution of HMXBs to the GW-detected population. 

Several studies have attempted to quantify the likelihood that an observed HMXB will evolve into a merging BCO. For instance, using the \textsc{compas} population synthesis code, \citet{Neijssel2021} evolved the binary system Cyg~X-1 based on the updated mass estimate of \citet{Miller2021}. Their results show that the system has about $\sim4$\% probability of merging as a BBH within a Hubble time, depending on the natal kick associated with the formation of the second BH. Similarly, using the same rapid population synthesis code, \citet{Romero-Shaw2023} found that roughly 8\% of BH-HMXBs are expected to merge as BBH or BHNS systems within a Hubble time, while their model predicts that Cyg X-1 does not satisfy the conditions for a future merger.
In contrast, using the \textsc{sevn} population synthesis code, \citet{Erika2025} showed that other specific configurations have much more promising fates. They investigated Cyg~X-3, the only Wolf-Rayet-compact object candidate observed in the Milky Way (MW), and found that about 70--100\% of Cyg X-3-like systems in a Wolf-Rayet-BH configuration are expected to become BCO progenitors.
However, even if some of these systems successfully merge, their properties present another challenge: the vast majority of BHs in HMXBs have masses below $<20 {\rm \ M_\odot}$, whereas a large fraction of detected BBHs have masses above this threshold. Moreover, spin measurements of BHs in X-ray binaries indicate a prevalence of large spins, seemingly at odds with measurements from GWs \citep[see e.g.,][]{Miller2015,Christopher2021}. 

Recent efforts have aimed at reconciling the apparent differences between the observed properties of HMXBs and those of merging BCOs inferred from GW observations. Some works suggest that these discrepancies may arise from selection effects rather than from fundamental physics differences between these two populations \citep{Fishbach2022,Liotine2023,Gallegos-Garcia2022}. For instance, using GWTC-2 data, \citet{Fishbach2022} find that BH masses in HMXBs are consistent with those in BBH mergers once GW selection biases and small-number statistics are taken into account. When BH spins are also included, only a small fraction of the BBH population appears HMXB-like, suggesting that $\sim$3\% of BBH mergers may originate from former HMXBs. Similarly, \citet{Liotine2023} study how the observational selection effects shape the detectability of HMXBs and merging BBHs. Their results indicate that  observable HMXBs and BBHs are formed at different epochs and distinct metallicity environments: while nearby HMXBs typically reside in Local Group galaxies with metallicities of $0.1-1$~Z$_\odot$, observable merging BBHs are expected to form at higher redshifts, where the metallicities are lower. In addition, the recent work by \citet{Misra2023} using {\sc posydon} binary population synthesis framework and employing an extensive sample of binary evolution models suggests that current discrepancies between observations and synthetic populations may reflect uncertainties in the underlying binary physics, rather than intrinsic differences in the observed systems. 
A complementary possibility is that the two populations largely form through different channels: a substantial fraction of BBH mergers may assemble dynamically in dense stellar environments, a pathway that never goes through an HMXB phase \citep[see e.g.,][]{Perna2019}.
In this dynamical scenario, hierarchical mergers are expected to produce high-mass BHs ($\gtrsim 50\,M_\odot$) with an isotropic spin distribution (\citealt{GWTC52026}; \citealt{Antonini2026}), putting an additional constraint on isolated HMXBs as the sole progenitors of BCOs.

Importantly, all the aforementioned studies usually assume homogeneous star formation and metallicity histories, rather than tracking the detailed evolution of individual galaxies. Assuming such homogeneous galaxy properties erases critical variations, such as metallicity gradients and age distributions, which can drastically affect the predicted binary evolution outcomes. In the context of BCO merger estimates, recent works have shown problems with relying on simplified analytical prescriptions. For example, \citet{Levina2026} compare analytical fits of the metallicity-dependent star formation rate density
against the IllustrisTNG 
simulations coupled with \textsc{compas}. 
They find that analytical models tend to overestimate the number of mergers at $z > 4$ and can lead to deviations in the BH mass distribution at $z < 2$. Likewise, the application of updated empirical models presents certain challenges. \citet{Sgalletta2025} (but see also \citealt{Boco2026}) show that the more accurately we model the metallicity-dependent cosmic star formation rate density evolution, the higher the binary black hole merger rate density that we estimate at redshift zero.

In this work, we focus on the population of \textit{isolated wind-fed} 
HMXBs in the MW, motivated by the fact that our Galaxy provides the most complete and well-characterized sample of such systems. Within the MW, the confirmed population of Galactic HMXBs is approximately 164 \footnote{\url{https://binary-revolution.github.io/HMXBwebcat/}}. Among these, 111 HMXBs possess a counterpart in the Gaia Early Data Release 3 \citep{Fortin2022,Fortin2023}. The MW therefore offers a unique constraining benchmark to assess theoretical models of HMXB formation and evolution. Our goal is to investigate the potential evolutionary link between present-day HMXBs and BCO mergers detectable as GW sources.  Although different classes of X-ray binaries exhibit distinct observational properties and physical behaviors, current population synthesis modeling cannot account for these specific observational features and requires a simplified and unified definition of HMXBs. Following previous studies \citep{Misra2023,Liotine2023,Gallegos-Garcia2022}, we define HMXBs as systems containing a compact object (BH or NS) and a massive stellar companion, without explicitly modeling accretion disk formation or imposing detailed observability criteria. A more detailed classification of X-ray binary sub-populations is beyond the scope of the present work.

We used a sample of 66 simulated MW-like galaxies from the TNG50 simulation \citep{Pillepich2023}, coupled with results from the population synthesis code {\sc sevn}. Within this self-consistent framework, we populate the simulated galaxies with HMXBs and analyze the resulting population in the context of the observational data, with particular emphasis on identifying systems that may evolve into merging BCOs. Briefly, the main updates to our population synthesis code, detailed in the subsequent sections, include using an NS mass distribution to account for more massive remnants and adopting a $\alpha_{\rm CE}$ value from \cite{Sgalletta2023} that best fits the MW neutron star merger rates. A systematic exploration of model uncertainties will be presented in a dedicated follow-up study.

The paper is structured as follows. In Section \ref{sec:methodology}, we describe the \textsc{sevn} population synthesis code, and the selection of MW-like galaxies from the TNG50 simulation. Section \ref{sec:populating} details the procedure for populating these simulated galaxies with HMXBs.
In Section \ref{sec:results}, we present the spatial distribution, radial profiles, and age demographics of the Galactic HMXB population. Section \ref{sec:discussion} explores the evolutionary pathways leading to binary compact objects, analyzing the physical drivers, such as supernova kicks and mass transfer efficiency, that determine the final merger fate. Finally, we summarize our main conclusions in Section \ref{sec:conclusions}.


\section{Methodology}\label{sec:methodology}
\subsection{ The population synthesis code {\sc sevn}}\label{sec:sevn_hmxb}

To simulate the Galactic HMXB population, we use the population synthesis code {\sc sevn} (Stellar EVolution \textit{N}-body; \citealt{Spera2015}; \citealt{Iorio2023}). This code computes single and binary stellar evolution by interpolating precomputed stellar tracks, such as {\sc parsec} (\citealt{Bressan2012}; \citealt{Nguyen2022}, \citealt{Costa2025}) and {\sc mist} (\citealt{Choi2016}). Binary interactions are handled via analytic and semi-analytic approaches. Key processes included in {\sc sevn} that are fundamental for the study of HMXBs and BCOs include wind-mass transfer, RLOF, common envelope (CE) evolution, stellar mergers or collisions, and orbital decay due to gravitational wave emission.

The comprehensive physical prescriptions and capabilities of the code are described in detail in \cite{Iorio2023} and \cite{Sgalletta2023}. In this study, we use a modified version of {\sc{sevn}}\footnote{Available in the official {\sc sevn} repository at \url{https://gitlab.com/sevncodes/sevn}}. Specifically, we introduce a modification to the standard prescriptions to better capture the physics of our systems of interest. 
The modification corresponds to the core-collapse supernova (SN) mechanisms. While {\sc sevn} includes the \textit{rapid} and \textit{delayed} models from \cite{Fryer2012}, we introduce an alternative Gaussian distribution for remnant masses in the \textit{rapid} model. The default \textit{rapid} model in {\sc sevn} selects NS masses using a single Gaussian distribution with $\mu = 1.33 \, \mathrm{M}_{\odot}$ and $\sigma = 0.09 \, \mathrm{M}_{\odot}$. For this work, we implement a double Gaussian distribution (Rapid-DG) with $\mu_1 = 1.36$ and $\mu_2 = 1.78$ ($\sigma_1 = 0.08$ and $\sigma_2 = 0.21$). This serves as a first approximation to reproduce the Galactic NS mass distribution presented in \cite{Abbott2023}. 

\subsubsection{Initial conditions and model parameters}\label{ref:IC}

Having defined the underlying physics and modifications, we initialize our binary systems. Primary zero-age main sequence (ZAMS) masses ($M_{1}$) are drawn from a Kroupa initial mass function (IMF; \citealt{Kroupa2001}):
\begin{equation}
    p(M_{\mathrm{ZAMS,1}}) \sim M_{\mathrm{ZAMS,1}}^{-2.3}, \quad M_{\mathrm{ZAMS,1}} \in [8,150] \, \mathrm{M}_{\odot}.
\end{equation}

The distributions for the secondary mass ($M_{2}$), orbital period ($P$), and eccentricity ($e$) are obtained from \cite{Sana2012}, derived from observations of O-type binary stars in nearby Galactic open stellar clusters:
\begin{equation}
    p(q) \sim q^{-0.1}, \quad q = \frac{M_{\mathrm{ZAMS,2}}}{M_{\mathrm{ZAMS,1}}}, \quad q \in [q_{\mathrm{min}}, 1.0],
\end{equation}
where $q_{\mathrm{min}} = \max \left( \frac{8.0}{M_{\mathrm{ZAMS,1}}}, 0.1 \right)$. For the period distribution, we adopt:
\begin{equation}
    p(\log P/\mathrm{days}) \sim (\log P/\mathrm{days})^{-0.55},
\end{equation} 
with $\log (P   /\mathrm{days}) \in [0.15, 5.5]$, and for the eccentricity:
\begin{equation}
    p(e) \sim e^{-0.42}, \quad e \in [0.0, 0.9].
\end{equation}

We assume an initial rotation equal to zero for all stars, representing the default model in {\sc sevn}. Tidal interactions are implemented following the model given by \citet{Hut1981}. For RLOF, orbital circularization takes place at the periastron at the onset of mass transfer. {\sc sevn} determines RLOF stability based on the critical mass ratio $q_{\mathrm{c}}$ taken from \cite{Hurley2002} using the formalism from \cite{Webbink1988} for giant donors, but always stable for main sequence (MS) and Hertzsprung-gap (HG) donors. While mass transfer from MS and HG donors is always stable, for donors in other phases it is stable only if the mass ratio $q = M_{\mathrm{d}}/M_{\mathrm{a}}$ (with $M_{\mathrm{d}}$ and $M_{\mathrm{a}}$ being the donor and accretor masses) satisfies $q < q_{\mathrm{c}}$. During RLOF, we set a conservative mass transfer by setting the mass-loss fraction to $f_{\mathrm{MT}} = 1.0$. For unstable mass transfer leading to CE evolution, we adopt the $\alpha_{\rm CE}-\lambda$ formalism \citep{Hurley2002}. For wind mass accretion rates we estimate the X-ray luminosity assuming stellar-wind accretion following the classical \cite{Bondi:194k} prescription under the fast-wind approximation (i.e., wind velocity exceeding the orbital velocity). Additionally, we impose an upper limit to the accreted mass of $0.8\ |\dot{M}_{d,\mathrm{wind}}|$, following Hurley (2002), to avoid unphysical accretion rates in eccentric systems. The efficiency parameter, which dictates the fraction of orbital energy converted into kinetic energy to unbind the envelope, is set to $\alpha_{\mathrm{CE}} = 3.0$. This value is adopted from \citet{Sgalletta2023} as the preferred parameter to reproduce the BNS merger rate in the MW. The envelope structure parameter is set to $\lambda = 1.0$, following the recipes in \citet{Claeys2014} as implemented in {\sc sevn} through the \texttt{BSE} option (see Appendix A1.4 in \citealt{Iorio2023}). Additionally, we assume that recombination energy is included in the estimate of the binding energy. Finally, natal kicks imparted to compact objects are drawn following \citet{Giacobbo2020}.
This parameter set was evaluated across 13 metallicities ($Z = 0.002$, $0.004$, $0.006$, $0.008$, $0.012$, $0.016$, $0.02$, $0.024$, $0.028$, $0.03$, $0.034$, $0.038$, and $0.04$). The lower bound was selected to probe the lowest $Z$ expected for stellar particles in MW-like galaxies in TNG50, while the upper bound is constrained by the maximum metallicity of the stellar tracks provided by the {\sc parsec} tables \citep{Costa2025}. Each simulation was initialized with $10^{7}$ binaries, yielding a total combined simulated mass of $M_{\mathrm{SEVN}} = 3.5 \times 10^{8} \, \mathrm{M}_{\odot}$. This corresponds to an effective total mass of $M_{\mathrm{pop}} = 4.07 \times 10^{9} \, \mathrm{M}_\odot$, assuming a binary fraction $f_{\mathrm{bin}} = 0.5$ \citep{Sana2012} and applying a correction factor (\(f_{\mathrm{corr}} = 0.172\)) for the incomplete IMF sampling due to our mass limits. Every binary system was then evolved until both stars formed compact remnants. A more extensive exploration of the model parameter space and its impact on the predicted HMXB population will be presented in a forthcoming paper (Vivanco Cádiz, \textit{in prep.}).

\subsubsection{HMXBs in {\sc sevn}}\label{sec:HMXB_in_SEVN}

We identify HMXBs in the simulated output by selecting systems composed of a compact object (BH or NS) and a massive companion star with $M \ge 8 \, \mathrm{M_{\odot}}$. The HMXB phase is defined as the period where the system's X-ray luminosity lies between $10^{31} - 10^{40} \, \mathrm{erg\, s^{-1}}$ \citep[motivated by][data catalog]{Fortin2023}. The intrinsic X-ray luminosity is calculated as:
\begin{equation}
    L_{\mathrm{X}, [0.5-10]\, \mathrm{keV}} = \mathrm{BC}(\lambda_{\mathrm{Edd}}) \times \epsilon \dot{M} c^{2} \quad \mathrm{erg\, s^{-1}},
\end{equation}
where $c$ is the speed of light, $\dot M$ is the mass-accretion rate via stellar winds, and $\mathrm{BC}(\lambda_{\mathrm{Edd}})$ is the bolometric correction as a function of the Eddington ratio. Following \cite{Anastasopoulou2022}, we convert the bolometric luminosity to the $0.5$--$10$ keV range using their correction factors.

The radiative efficiency, $\epsilon$, is treated separately for BHs and NSs. For BHs, we adopt the prescription from \cite{Podsiadlowski2003}:
\begin{equation}
    \epsilon = 1 - \sqrt{1-\left(\frac{M_{\mathrm{BH}}}{3M_{\mathrm{BH}}^{0}}\right)^{2}}, \quad \text{for} \quad M_{\mathrm{BH}} < \sqrt{6}M_{\mathrm{BH}}^{0},
\end{equation}
where $M_{\mathrm{BH}}$ and $M_{\mathrm{BH}}^{0}$ are the current BH mass and the BH mass at formation, respectively. For NSs, the efficiency is calculated as:
\begin{equation}
    \epsilon = \frac{GM_{\mathrm{NS}}}{R_{\mathrm{NS}}c^{2}},
\end{equation}
where $G$ is the universal gravitational constant, and $M_{\mathrm{NS}}$ and $R_{\mathrm{NS}}$ are the mass and radius of the NS. Additionally, we adopt a maximum Eddington accretion ratio of 20. Although a detailed analysis of Ultraluminous X-ray Sources is beyond the scope of this study, this modification allows for super-Eddington accretion rates, which is necessary to reproduce the high-luminosity tail of the distribution and to prevent an artificial pile-up of sources at the Eddington limit.

To extract the population of HMXBs from {\sc sevn}, we randomly select systems with a probability weighted by the time-steps relative to the total duration of the HMXB phase ($w = \Delta t / \sum \Delta t$). This weighting accounts for {\sc sevn}’s adaptive time-step scheme, where the output resolution increases when system properties vary by more than a specific threshold (set as the default value of 5\%). As a result, the selection probability scales directly with the physical phase duration, forcing that longer-lived evolutionary states (wind accretion vs. CE/RLOF phases, prior to filtering exclusively for wind-fed systems) are 
sampled proportionally.

In Appendix~\ref{ap:efficiencies}, we study the link between the population of HMXBs in {\sc sevn} and the merger efficiency using the fiducial model for binary compact objects. In the next section, we present the method implemented for combining the population of HMXBs from {\sc sevn} with the MW-like galaxy sample from the IllustrisTNG50 simulation.

\subsection{IllustrisTNG}\label{sec:TNG50_MW}

IllustrisTNG is a suite of magnetohydrodynamical cosmological simulations that model the formation and evolution of galaxies under the standard $\Lambda$CDM paradigm \citep{Pillepich2018,Nelson2019,Marinacci2018,Springel2018}. The simulations are performed with the moving-mesh code {\sc arepo} \citep{Springel2010} and include subgrid models for star formation, stellar evolution, chemical enrichment, primordial and metal-line cooling of the gas, stellar feedback from type II and Ia supernovae and from AGB stars, as well as supermassive BH seeding and growth, and multimode feedback among other characteristics.

Here we use a subset of galaxies from the IllustrisTNG50-1 (hereafter TNG50, \citealt{Pillepich2019,Nelson2019b}). TNG50 represents a simulated box of (51.7~cMpc)$^3$ with the highest resolution available in the TNG suite (\(4.5 \times 10^{5}\ \mathrm{M}_{\odot}\) for dark matter particles and \(8.5 \times 10^{4}\ \mathrm{M}_{\odot}\) for the initial gas mass cells). To generate the MW's HMXB population, we used a subset of  MW- and Andromeda-like (MW/M31-like) galaxies from \citet{Pillepich2024}. We outline the selection process in the following section. 

\subsubsection{MW-like galaxies in TNG50}\label{ref:MW-like}

As mentioned above, we utilize a subset of MW/M31-like galaxies from the catalog reported by \cite{Pillepich2024}. This catalog includes 198 galaxies, from which we selected 66 based on the following criteria:

\begin{enumerate}
    \item Stellar Mass: We selected galaxies with stellar masses within $R < 30$ kpc ranging from $10^{10.5}$ to $10^{10.9} \,{M}_{\odot}$ to focus specifically on MW analogs.
    \item Stellar Particle Requirements: Selected galaxies must contain at least $10^{3}$ stellar particles satisfying these conditions:
    \begin{itemize}
        \item Stellar Age: Particles must be younger than 40 Myr. This threshold is informed by {\sc sevn} simulations, which show that HMXB systems do not persist beyond $\sim 35$ Myr after formation.
        \item Distance: Particles must be within 65 kpc of the galactic center to exclude satellite or background particles present in the large-scale TNG50 datacubes.
        \item Metallicity: Particles must have a metallicity between $Z = 0.002$ and $Z = 0.04$, consistent with the range established in the {\sc sevn} lookup tables.
    \end{itemize} 
    \item Visual Inspection: Two galaxies (IDs 487742 and 552581) were discarded after visual inspection. The former was in the process of merging with a satellite galaxy, while the latter lacked the characteristic spiral morphology.
\end{enumerate}

The resulting sample of 66 MW-like galaxies spans star formation rates (SFR) from $\sim 1.7$ to $12 \, {M}_{\odot} \, \mathrm{yr}^{-1}$ and gas metallicities between 0.014 and 0.036. For comparison, observational estimates for the MW place its stellar mass in the range ${\rm M}_{\star} \sim (4-7.3) \times 10^{10} \, {M}_{\odot}$  and its SFR at $1.65 \pm 0.19 \, { M}_{\odot} \, \mathrm{yr}^{-1}$ \citep{Bland-Hawthorn2016}.
In the next section, we describe the procedure used to populate the simulated MW-like galaxies from TNG50 with HMXB systems simulated with {\sc sevn}.

\begin{figure*}[t]
    \centering
    \includegraphics[width=\textwidth]{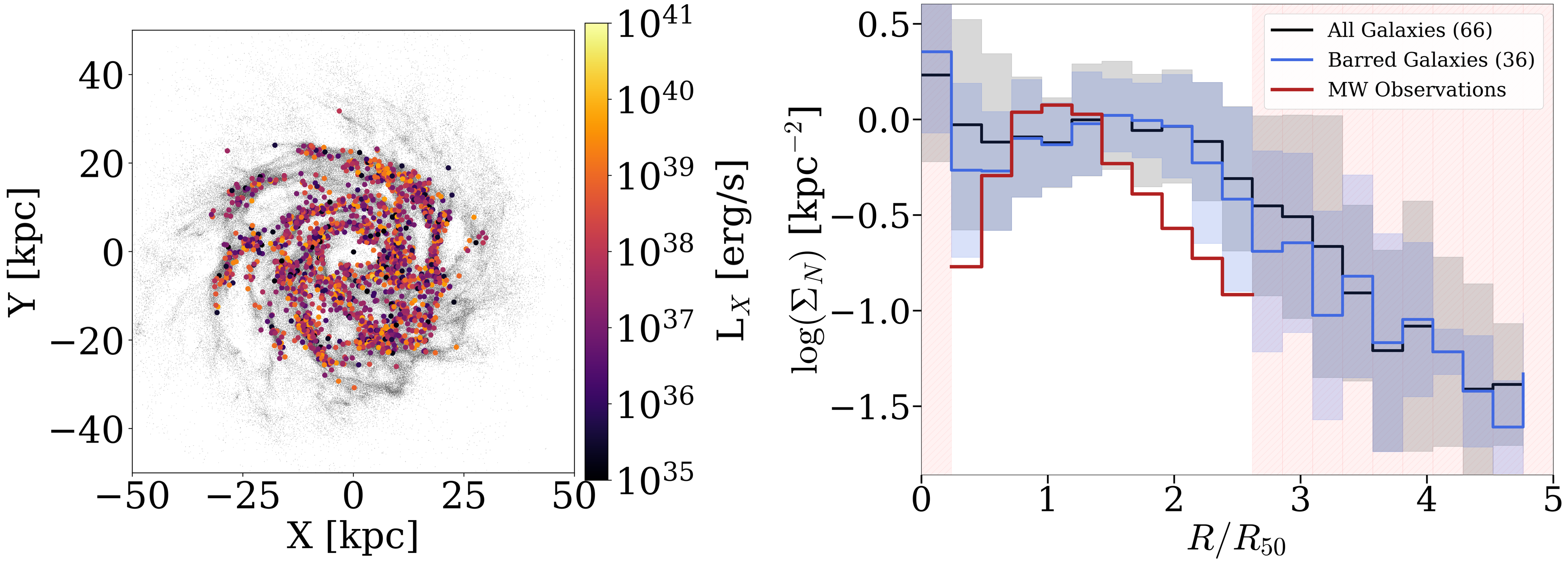}
    \caption{
    \textit{Left panel:} Gas surface density map of a representative MW-like galaxy (ID = 372755), overlaid with the spatial distribution of stellar particles hosting HMXBs, color-coded by X-ray luminosity. 
    \textit{Right panel:} Radial surface number density profile of HMXBs as a function of the normalized galactocentric radius ($R/R_{50}$). The median profiles for the full simulated sample (black) and the barred galaxy subsample (blue) are compared against observational data from \cite{Fortin2023} (red). Grey and blue shaded regions indicate the median $\pm$ median absolute deviation for the simulated samples, while the light red shaded areas indicate regions lacking observations. For this analysis, we use the observed-like HMXB sample ($10^{35}-10^{40} \ {\rm erg\ s^{-1}}$).}
    
    \label{fig:mw_properties}
\end{figure*}

\subsection{Populating MW-like galaxies with HMXBs}\label{sec:populating}

We populate the simulated galaxies by assigning HMXB systems to the stellar particles based on their fundamental properties: mass ($M_{\mathrm{sp}}$), age ($t_{\mathrm{sp}}$), and metallicity ($Z_{\mathrm{sp}}$). Our method follows earlier studies \citep[see, for instance,][]{Artale2019a}, and incorporates several updates. The procedure for this assignment is as follows:

\begin{enumerate}
    \item Based on the metallicity $Z_{\mathrm{sp}}$ of the stellar particle, we select the corresponding {\sc sevn}-simulated HMXB catalog with the closest available metallicity. This procedure introduces a minor mismatch, with a median relative difference of \(4.2\times10^{-2}\) (approximately  \(\sim 4\%\)). Overall, \(\sim90\%\) of the systems show relative differences below 0.10, while the most extreme cases (\(99\%\)) remain within a \(20\%\) relative to the SEVN metallicity values.
    \item Using the stellar particle age $t_{\mathrm{sp}}$, we determine the number of unique HMXB systems ($N_{\mathrm{SEVN}}$) in the selected {\sc sevn} catalog that fall within the interval $t_{\mathrm{sp}} \pm 1$ Myr. This time window accounts for the temporal resolution of TNG50 at $z=0$, which is approximately $\pm 2$ Myr.
     \item We compute the expected number of HMXB systems ($\theta$) associated with the stellar particle using the relation:
    \begin{equation}
        \theta = \frac{N_{\mathrm{SEVN}} \times M_{\mathrm{sp}}}{M_{\mathrm{pop}}},
    \end{equation}
    \item The final count of HMXBs assigned to the stellar particle, $N_{\mathrm{HMXBs}}$, is drawn from a Poisson distribution:
    \begin{equation}
        N_{\mathrm{HMXBs}} \sim \text{Poisson}(\theta).
    \end{equation}
    which provides an integer count of HMXBs associated with the stellar particle.
    \item Finally, we randomly select $N_{\mathrm{HMXBs}}$ systems from  the filtered subset of the {\sc sevn}-simulated HMXB catalog. 
\end{enumerate}

This methodology allows us to create a comprehensive population of HMXBs that preserves their dynamical properties (such as period, eccentricity, masses, and natal kicks) and enables the analysis of their spatial distribution within simulated MW-like galaxies. In the following, we define the observed-like HMXB sample as those sources with luminosities between $10^{35} - 10^{40} \ {\rm erg\ s^{-1}}$ \citep[see,][]{Fornasini2023,Fortin2023}. We use the term full sample to denote all sources within the luminosity range $10^{31}-10^{40} \ {\rm erg\ s^{-1}}$.

\section{Results}\label{sec:results}
\subsection{HMXB populations in MW-like galaxies}\label{HMXBs_population}

\begin{figure*}[h]
    \centering
    \includegraphics[width=\linewidth]{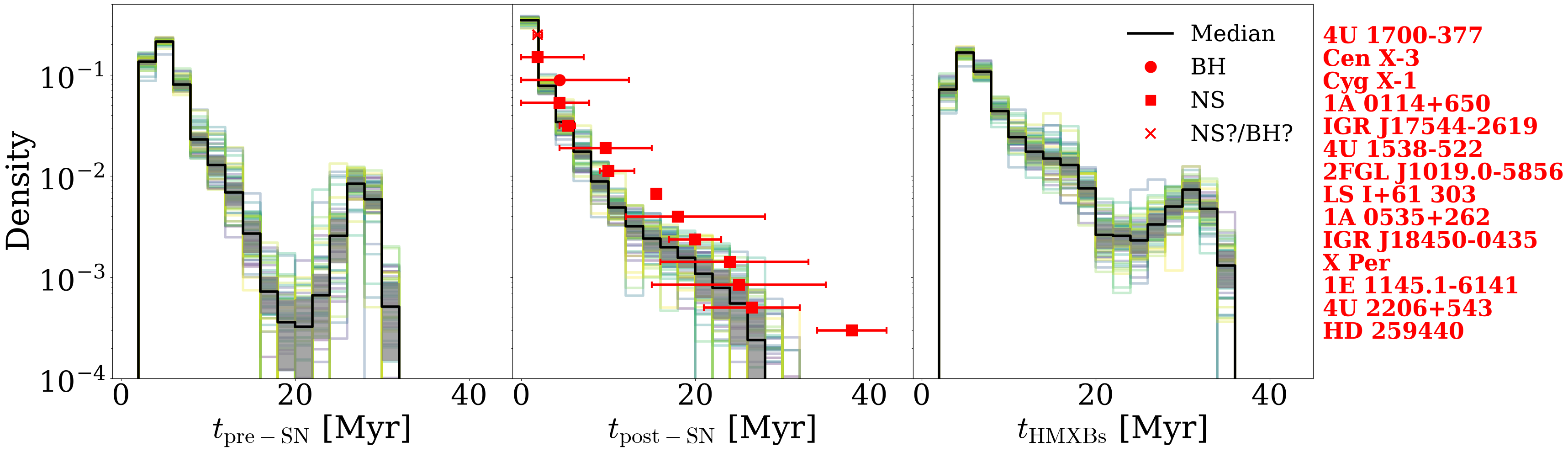}
    \caption{Normalized age distributions of HMXBs from the MW-like galaxy sample. The panels display the pre-SN age, ($t_{\mathrm{pre-SN}}$, left), post-SN age, ($t_{\mathrm{post-SN}}$, middle), and the HMXB age  computed as the sum of the pre-SN and post-SN age (right, $t_{\mathrm{HMXB}}=t_{\mathrm{pre-SN}}+t_{\mathrm{pre-SN}}$). Faint colored lines represent individual simulated galaxies, while the solid black line represents their median. The shaded area indicates the median absolute deviation. Red markers in the middle panel show observed post-SN ages from \citet{Fortin2022}. Marker shapes denote the compact object type: squares for NS, circles for BH, and crosses for unconfirmed candidates. These observational points are vertically offset in logarithmic space for visual clarity and \textit{do not} map to the y-axis values. Source names are listed on the far right, corresponding to the markers from top to bottom. We exclude SS 433, available in \citet{Fortin2022}, as its age post-SN is an upper limit ($< 60$~Myr).}
    \label{fig:age_distribution}
\end{figure*}

The left panel of Fig. \ref{fig:mw_properties} illustrates the spatial distribution of the HMXB population in a representative MW-like galaxy (ID = 372755), used to identify the physical configuration that best reproduces observational data. Each circular point represents a stellar particle of the selected galaxy in the TNG50 simulation, color-coded by its total X-ray luminosity, computed as the sum of all HMXB systems hosted within that particle. As expected, the HMXB population closely traces the galactic spiral arms. This spatial correlation is a direct consequence of our selection criteria, which focus on young stellar populations ($< 40$~Myr; see right panel of Fig. \ref{fig:age_distribution}). Since HMXB formation is intrinsically linked to the age of the stellar particles, modifying this cutoff would not significantly alter the observed spatial distribution; specifically, stellar particles older than $\sim 35$~Myr do not host HMXB systems in our model.

The right panel of Fig.~\ref{fig:mw_properties} displays the radial surface number density profile for HMXBs with luminosities in the range $10^{35} \text{--} 10^{40} \text{ erg s}^{-1}$ (observed-like sample), as a function of the normalized galactocentric radius ($R/R_{50}$), where $R_{50}$ is the half-light radius. We derived these numerical profiles for the full simulated sample and a subsample of barred galaxies (selected following \citealt{Rosas-Guevara2022}) and compared them against the observational Galactic HMXB distribution from \cite{Fortin2023}. For a consistent comparison, both simulated and observed radial distances were normalized using a half-light radius of $R_{50} = 5.75 \pm 0.38 \text{ kpc}$, as adopted from \cite{Lian2024} and the
radial profile was divided by \(A_{\mathrm{obs}} = 1/3\) accounting for the area coverage of the observed HMXBs positions. We note that while MW-like galaxies in the TNG50 simulation are, on average, larger than the present-day MW, and their size growth is found to be relatively strong at late cosmic times ($z \lesssim 0.7$) in contrast with observational constraints \citep{Hasheminia2022}, the normalization by $R_{50}$ allows us to mitigate potential biases.

The normalized age distribution of the simulated HMXB population is shown in Fig. \ref{fig:age_distribution}. We split the HMXB age into two components: the pre-SN age ($t_{\mathrm{pre-SN}}$), spanning from the formation of the binary system to the first SN event (left panel), and the post-SN age ($t_{\mathrm{post-SN}}$), covering the interval from the first SN event to the onset of the HMXB phase (middle panel). The HMXB age  is obtained by adding these two contributions (i.e., $t_{\mathrm{HMXBs}} = t_{\mathrm{pre-SN}} + t_{\mathrm{post-SN}}$, see right panel).

The figure presents the age distributions for each individual galaxy, along with the corresponding median distributions. The findings reveal no substantial differences among various MW-like galaxies, suggesting that the underlying stellar population model is the primary driver of these demographic trends.

We compare our post-SN ages with the kinematical ages derived by \citet{Fortin2022}, who trace Galactic HMXBs back to their birthplaces through orbit integration and define the kinematical age as the time elapsed since the first SN. In our simulations, HMXBs tend to have short post-SN ages, a direct consequence of the short lifetimes of their massive donors, which restrict the HMXB phase to a brief window after the first SN. The predicted ages are broadly consistent with the observed ones, although the observed sample does not display the same preference for short post-SN ages. This comparison should nevertheless be taken with caution: the observational sample is small, and the kinematical ages carry large and often asymmetric uncertainties, with several systems constrained only by upper limits.

A more detailed analysis of the population properties according to age is presented in Appendix \ref{appendix:b2}. 
Our main findings are that HMXB systems younger than 10 Myr mainly come from progenitors more massive than $\sim 20~{M_\odot}$ and are predominantly BH-HMXBs. In contrast, systems older than 10 Myr largely originate from progenitors with initial masses of about $8$–$25~{M_\odot}$ and are dominated by NS-HMXBs. In addition, a small fraction of BH-HMXBs exceeds 10 Myr and tends to have relatively low-mass companions ($<20~{M_\odot}$). Most NS-HMXBs, on the other hand, generally have ages $>10$ Myr and are linked to companion stars with masses $\lesssim 25~ {M_\odot}$.
Furthermore, we find that X-ray luminosity depends on system age and compact object type. Young BH-HMXBs ($<10$ Myr) span a wide luminosity range ($10^{35}-10^{40} {\rm erg\ s^{-1}}$), while older BH-HMXBs are typically dimmer. Young NS-HMXBs peak at a higher luminosity ($\sim 10^{38} {\rm erg \ s}^{-1}$) than older NS-HMXBs ($\sim 2\times10^{37} {\rm erg\ s}^{-1}$).

\subsection{HMXBs as Progenitors of BCOs}

\begin{figure*}[h!]
    \centering
    \includegraphics[width=\textwidth]{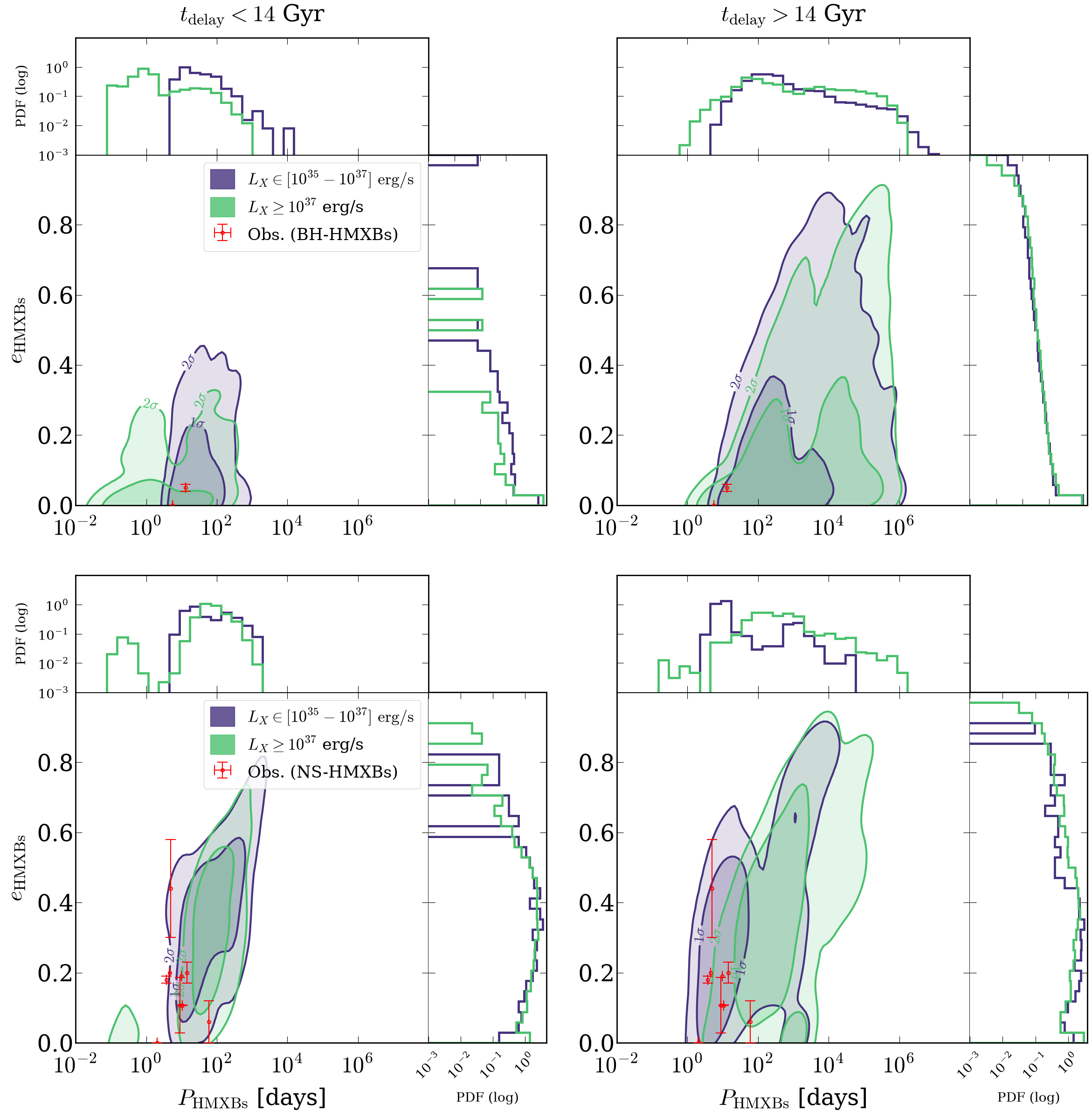}
    \caption{Eccentricity as a function of the orbital period for simulated and observed BH-HMXBs (top panel) and NS-HMXBs (bottom panel). In each panel, we show $1\sigma$ and $2\sigma$ ($68\%$ and $95\%$) KDE contours for the simulated HMXB population that will evolve into BCOs, compared with observational data (points with error bars) from \citet{Fortin2023}. The left column shows merging systems ($t_{\mathrm{delay}} <$14 Gyr), while the right column shows non-merging systems within a Hubble time ($t_{\mathrm{delay}} >$14 Gyr). We split the population by X-ray luminosity into
    bright ($L_X \ge 10^{37}~{\rm erg/s}$) and faint systems ($L_X \in 10^{35} - 10^{37}~{\rm erg/s}$). In each panel, the marginal histograms illustrate how eccentricities and periods are distributed, separated into bright and faint HMXB systems.}
    \label{fig:period_eccentricity_hmxbs}
\end{figure*}

\begin{figure*}[h!]
    \centering
    \includegraphics[width=\textwidth]{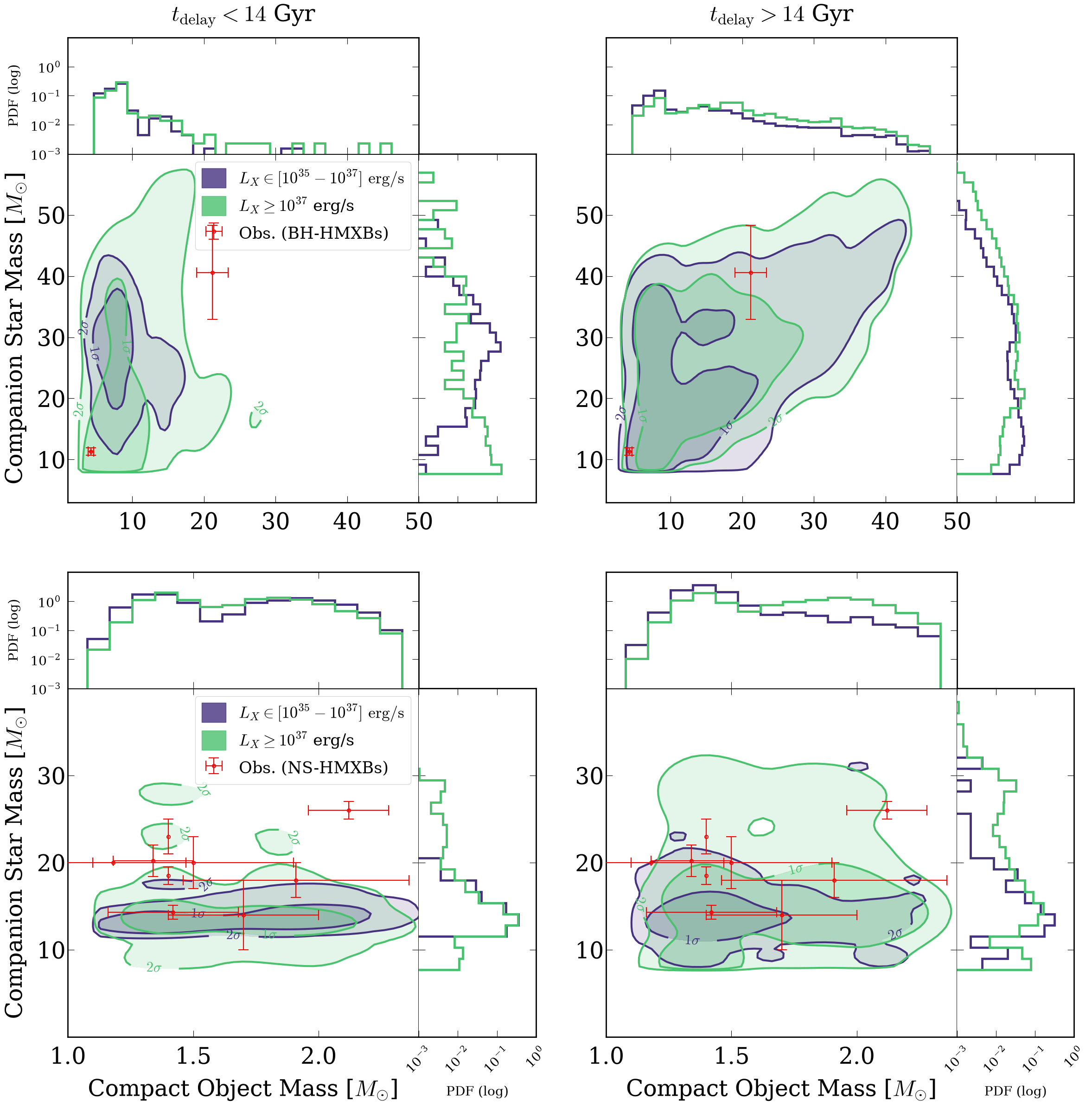}
    \caption{Comparison of the companion star mass versus compact object mass for simulated and observed BH-HMXBs (top panel) and NS-HMXBs (bottom panel). In each panel, we show $1\sigma$ and $2\sigma$ ($68\%$ and $95\%$) KDE contours for the simulated HMXB population that will evolve into BCOs, compared with observational data (points with error bars) from \citet{Fortin2023}. The left column shows merging systems ($t_{\mathrm{delay}} <$14 Gyr), while the right column shows non-merging systems ($t_{\mathrm{delay}} >$14 Gyr). We further divide the population by X-ray luminosity into
    bright ($L_X  \ge 10^{37}~{\rm erg/s}$) and faint systems ($L_X \in [10^{35} - 10^{37}]~{\rm erg/s}$). Each panel shows marginal histograms of the companion and compact object masses, split into bright and faint systems.}
    \label{fig:mco_vs_m_com_bh_ns_hmxbs}
\end{figure*}

\begin{table*}[t]
\renewcommand{\arraystretch}{1.5} 
\centering
\caption{Census of HMXB populations per MW-like galaxy. The columns compare the population restricted by X-ray luminosity ($10^{35} \leq L_X \leq 10^{40}$\,erg\,s$^{-1}$) versus the complete population (No $L_X$ cut). For each category, the top row indicates the total number of systems, while the bottom row shows the contribution from BCO mergers with delay times $t_{\rm delay} < 14$\,Gyr. Subpopulations are listed as (BHBH/BHNS) for the BH-HMXB columns and as (NSNS/NSBH) for the NS-HMXB columns. All values are reported as the median of the simulated sample, with uncertainties corresponding to the 16th and 84th percentiles.}
\label{table:fractions}
    \begin{tabular}{c|c|c|c}
    \hline\hline
    \multicolumn{2}{c|}{ Observed-like ($10^{35}-10^{40}$ erg/s)} & \multicolumn{2}{c}{Full Sample} \\
    \cline{1-2}\cline{3-4}
    BH-HMXB & NS-HMXB & BH-HMXB & NS-HMXB \\ 
    (BHBH / BHNS) & (NSNS / NSBH) & (BHBH / BHNS) & (NSNS / NSBH) \\ \hline

    $1633^{+1136}_{-724}$ & $358^{+268}_{-112}$ & $3024^{+2139}_{-1317}$ & $3282^{+2474}_{-1115}$ \\
    $(15^{+21}_{-11}) / (7^{+6}_{-4})$ & $(35^{+23}_{-13}) / (2^{+4}_{-2})$ & $(16^{+23}_{-12}) / (9^{+7}_{-5})$ & $(274^{+222}_{-70}) / (5^{+7}_{-4})$ \\ \hline

    \end{tabular}
\end{table*}

We now examine the subsequent evolution of the HMXB population to determine which systems are capable of producing merging BCOs. The evolutionary trajectory of these binaries is governed by several critical physical processes, including mass-transfer efficiency, stellar wind mass-loss rates, CE evolution, and supernova natal kicks, all of which determine whether a system will remain gravitationally bound or be disrupted.

A comprehensive analysis of the HMXB population and their future compact object remnants can be performed by examining their orbital properties in the period-eccentricity ($P_{\rm HMXBs}\text{--}e_{\rm HMXBs}$) plane. 
Figure \ref{fig:period_eccentricity_hmxbs} presents these distributions for BH-HMXBs and NS-HMXBs, overlaid with observational data from \citet{Fortin2023}. 
Each figure is split in two columns corresponding to merger or non-merger given by their delay times ($t_{\text{delay}}$), defined as the sum of the evolutionary time until the formation of the BCO and the subsequent gravitational wave inspiral time \citep[see,][for further details]{Iorio2023}. We define HMXBs that end as BCO mergers as those whose delay time is shorter than the Hubble time, i.e., with $t_{\text{delay}} < 14$~Gyr, while systems with $t_{\text{delay}} > 14$~Gyr are classified as non-mergers.

Our results show that the progenitor regions for future GW events are distinct for each population. For BH-HMXBs, potential merger candidates are primarily confined to orbital periods $P_{\rm HMXBs} \lesssim 10^2$ days and low to moderate eccentricities ($e_{\rm HMXBs} \lesssim 0.4$). In contrast, the NS-HMXB merger progenitor region extends to longer periods ($P_{\rm HMXBs} \lesssim 10^3$ days) and encompasses a significantly broader range of eccentricities, reaching up to $e_{\rm HMXBs} \approx 0.9$.
Furthermore, for BH-HMXBs merging within a Hubble time, bright systems ($L_X \ge 10^{37}~{\rm erg/s}$)  tend to show lower periods and eccentricities than faint systems ($L_X \in [10^{35} - 10^{37}]~{\rm erg/s}$). Similar trends are also found for the period of NS-HMXBs. This would be a consequence of tight orbits producing brighter systems.

 From Figure~\ref{fig:period_eccentricity_hmxbs}, we find that the HMXB observed sample exhibits orbital periods and eccentricities that fall within a region where merging and non‐merging systems overlap, suggesting that these two parameters alone are insufficient to determine their eventual outcome as BCO mergers.

We further investigate the distribution of companion star masses versus compact object masses in Figure \ref{fig:mco_vs_m_com_bh_ns_hmxbs}. For BH-HMXBs, the merging population is characterized by a wide range of companion masses, between $\sim 8-50~M_\odot$, whereas in the case of NS-HMXBs, the companion masses are concentrated around $\sim 8-30~M_\odot$, corresponding to those systems that remain bound after the second supernova event. Furthermore, for BH-HMXBs, we find that bright sources show a large spread for the companion star masses but peaking near $\sim 8~M_\odot$, while faint ones present a peak around $\sim 30~M_\odot$. Taken together with the results from Figure~\ref{fig:period_eccentricity_hmxbs}, suggest that bright systems generally have shorter orbital periods. However, for such systems to both endure at these short periods and eventually form BCO mergers, the companion star must typically be a less massive star.

The contours illustrate that the simulated populations successfully encompass the majority of the observed Galactic systems. Overall, our findings indicate that several of the observed HMXBs reside in regions where they could, in principle, form a BCO merger (provided they survive the second supernova and retain a sufficiently short orbital period to coalesce within a Hubble time) since the parameter spaces for merging and non-merging systems overlap.

To quantify the impact of observational selection effects, Table \ref{table:fractions} presents a census of the HMXB population per MW-like galaxy. We note that the associated uncertainties are substantial, reflecting the stochastic nature of HMXB populations in individual MW–like galaxies. We also find that imposing an $L_X$ threshold on the HMXB population drastically reduces the number of NS-HMXBs, decreasing their population by a factor of $\sim 4$ compared to the BH-HMXBs.

We find that approximately $\sim 0.2-3.2\%$ of BH-HMXBs eventually form BCO mergers (BBH and BHNS), whereas for NS-HMXBs this fraction is $\sim 3.6-23.4\%$. 
If we restrict the population by X-ray luminosities in the range $10^{35} \le L_X \le 10^{40}$~erg s$^{-1}$, we obtain that about $\sim 0.3-5.4\%$ of BH-HMXBs produce BBH or BHNS mergers, and $\sim 3.5-26.0\%$ of NS-HMXBs evolve into BCO mergers, including BNS and NSBH systems. 

\section{Discussion}\label{sec:discussion}

In this section, we interpret our main results on the connection between Galactic HMXBs and merging BCOs. We focus on the physical factors that shape the merger progenitor population: the pre-HMXB binary-interaction history and subsequent interaction dynamics during the HMXB phase. We then discuss how these mechanisms differ between NS and BH-hosting systems and identify the parameters that primarily determine whether a system merges within a Hubble time. Finally, we summarize the main observational limitations and modeling uncertainties affecting the comparison with the MW sample.
\subsection{Evolutionary Pathways: The Pre-HMXB Phase}\label{pre_HMXBs}
HMXBs represent the subset of massive binaries that survive their first supernova while remaining gravitationally bound. Their present-day orbital configurations therefore, encode the sequence of binary interactions that occurred prior to (and around) compact-object formation, such as stable mass transfer (SMT)  and CE evolution.

To interpret which observed-like HMXBs are expected to evolve into merging BCOs, we summarize the pre-HMXBs evolutionary history using the channel classification of \citet{Iorio2023} \citep[see also,][]{Broekgaarden2021}. For clarity, we group these pathways into SMT-dominated channels (CI--CII, where CI systems that later experience at least one CE phase), early CE channels (CIII-CIV), and non-interacting channels (C0 and CV). The details of the formal channel definitions are provided in Appendix~\ref{appendix:channels}.
Fig.~\ref{fig:dominant_channels_p_e} shows how these dominant channels populate the orbital period–eccentricity plane, categorized by compact object type and $t_{\mathrm{delay}}$. Instead of showing individual systems, we display the dominant formation channel in each bin, while the marginal histograms on the axes quantify the normalized contribution of each channel to the total population. The distribution reveals a clear separation between the progenitors of merging and non-merging systems, shaped by their evolutionary history prior to the HMXB phase.

\begin{figure*}[t]
    \centering
    \includegraphics[width=\textwidth]{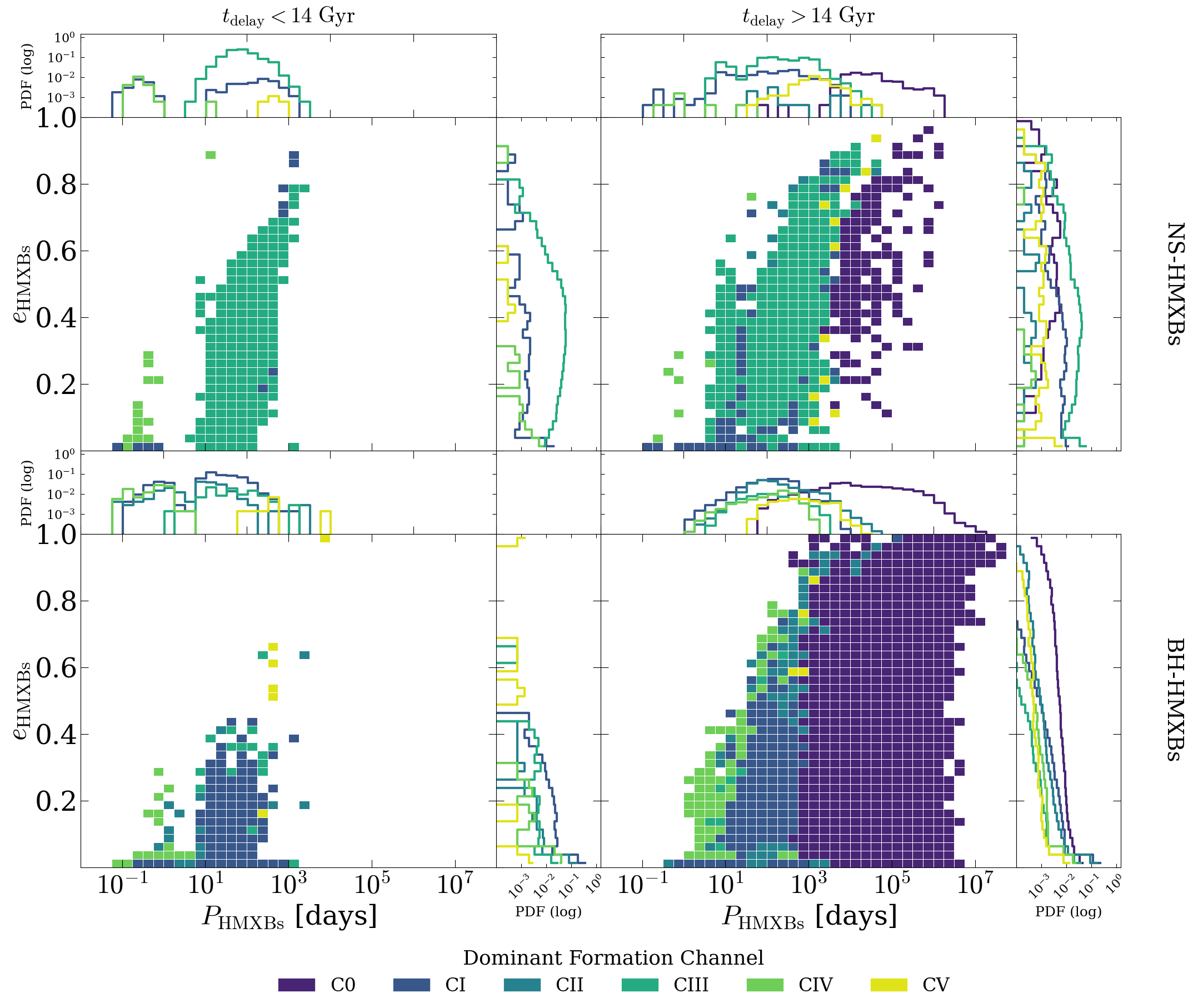}
    \caption{The dominant formation channels for HMXB progenitors in the orbital period versus eccentricity plane. Panels are organized by compact object type (rows) and merger delay time (columns). The left column shows merging systems ($t_{\mathrm{delay}} < 14$\,Gyr), while the right column shows non-merging systems ($t_{\mathrm{delay}} > 14$\,Gyr). Marginal histograms on the top and right axes display the normalized distribution of orbital periods and eccentricities, respectively, for each channel. Colors indicate the evolutionary channels following \citet{Iorio2023}: CI (blue) and CII (teal) represent SMT, while CIII (green) and CIV (light green) denote CE evolution. The non-merging population is characterized by a prevalence of C0 (violet) and Channel V (yellow), highlighting that systems avoiding strong orbital contraction tend to remain in wide orbits.}
    \label{fig:dominant_channels_p_e}
\end{figure*}

Future mergers ($t_{\mathrm{delay}} < 14$~Gyr) are driven by systems that have undergone significant orbital hardening. Specifically, the marginal histograms for merging systems (left panels) show two distinct peaks at short periods ($P_{\mathrm{HMXBs}} \sim 1$ and $10^2$ days), dominated by channels CIII and IV (green shades), where early CE evolution efficiently extracts angular momentum. However, Fig.~\ref{fig:dominant_channels_p_e} highlights that binary interaction is a necessary but not sufficient condition for merger. While the merging population is dominated by interacting channels, the non-merging population (right panels) also includes systems from CII, III, and IV. In many cases, SMT leads to orbital widening, stranding binaries in wide configurations ($P_{\mathrm{HMXBs}} > 100$~days) that prevent a merger within a Hubble time.

We also find a distinct separation based on the nature of the compact object. The NS population is largely dominated by formation channels involving a CE phase (CIII and IV), especially among merging systems. This suggests that most NS-HMXB progenitors require significant orbital contraction to survive strong NS natal kicks and eventually merge. In contrast, the BH population shows a higher frequency of SMT (CI and CII). Notably, non-merging BH systems are dominated by Channel II. As BH progenitors are more massive, they tend to undergo stable mass exchange rather than entering a CE phase; while this stability ensures binary survival, it often results in wide orbits that prevent future coalescence.
In summary, pre-HMXB evolution acts as a primary filter, segregating the population into potential merger candidates (compact, post-CE systems) and those destined to remain wide binaries (post-SMT or non-interacting systems).

\subsection{The HMXB phase: interaction dynamics and orbital evolution}

While formation channels summarize the pre-remnant history, the HMXB phase is shaped by the mass and angular-momentum exchange that takes place after the first supernova, through wind accretion, RLOF, or unstable CE evolution.
These interactions set the orbital response of the binary and therefore the initial conditions for the second supernova, which ultimately determine whether the system survives to form a BCO and, if so, whether it can merge within a Hubble time.
In this work, we classify the simulated HMXB population into four distinct interaction categories (dominant formation subchannels): (i) no further interaction, (ii) SMT only, (iii) a single CE episode or (iv) multiple CE episodes. 

\begin{figure*}
    \centering
    \includegraphics[width=\textwidth]{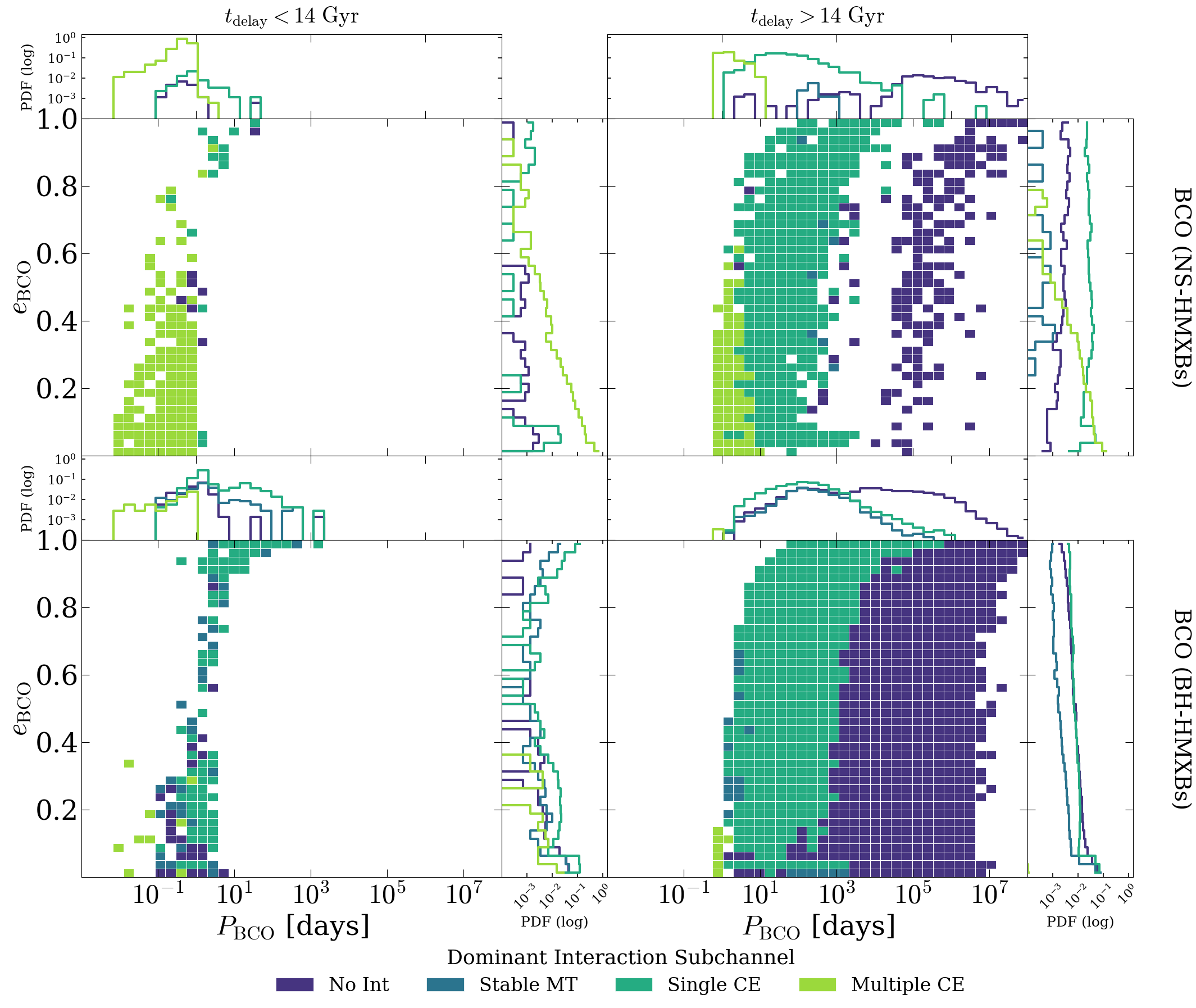}
    \caption{The dominant subformation channels for BCOs originating from NS-HMXBs (top row) and BH-HMXBs (bottom row) during the HMXBs phase. The columns distinguish between systems that merge within a Hubble time ($t_{\mathrm{delay}} < 14$\,Gyr, left) and those that do not (right). Marginal histograms on the top and right axes display the normalized distribution of orbital periods and eccentricities, respectively, for each subchannel. Colors indicate the evolutionary subchannels as No Interaction (violet), SMT (blue), Single CE (teal), and Multiple CE (yellow).}
    \label{fig:interaction_dynamics}
\end{figure*}

Figure \ref{fig:interaction_dynamics} illustrates the orbital period and eccentricity distribution for BCOs originating from NS-HMXB and BH-HMXB, color-coded by the dominant interaction sub-channel. The marginal histograms allow us to distinctly identify the prevalence of each interaction type.
For the merging BCO population derived from NS-HMXBs (top-left panel), most systems are concentrated at short orbital periods ($P_{\rm BCOs} < 1$ day). These systems are overwhelmingly dominated by the 'Multiple CE' sub-channel (yellow in the marginals), reaching minimum values of $\sim 10^{-2}$ days. A notable exception is the subset of high-eccentricity systems ($e_{\rm BCOs} \sim 0.9$), which extend to periods of nearly $10^2$ days. Evolutionary analysis indicates that most NS-HMXBs typically underwent a CE phase prior to the first supernova and subsequently survived multiple CE events after the explosion. This repeated unstable interaction effectively hardens the orbit, allowing the system to merge within a Hubble time.

Similarly, merging BCOs from BH-HMXBs (bottom-left panel) are centered around $P_{\rm BCOs} \sim 10$ days. Here, the marginal distributions reveal a more complex mix, with contributions from both CE and SMT sub-channels.
We again observe a distinct population of high-eccentricity systems ($e_{\rm BCOs} \approx 1$) at significantly wider periods ($\sim 10^3$ days),  suggesting that under specific high-eccentricity conditions induced by natal kicks, even systems that have undergone stable interactions can eventually merge.

Finally, the non-merging population (right panels) consists of a diverse mix of systems from both HMXB types. Although these binaries span all formation channels and subchannels—including those that experienced significant interaction—the orbital contraction was insufficient to drive them into the gravitational-wave emitting regime within 14 Gyr.

\subsection{Physical drivers of merger vs non-merger systems}

\subsubsection{The role of supernova kicks}

To understand why some systems fail to merge despite undergoing multiple interaction phases, we analyze the CDFs of the second supernova natal kick magnitude ($v_{\mathrm{kick}}$), orbital eccentricity, and orbital period for both BNS and BBH populations. 

\begin{figure*}[t]
    \centering
    \includegraphics[width=1.\textwidth]{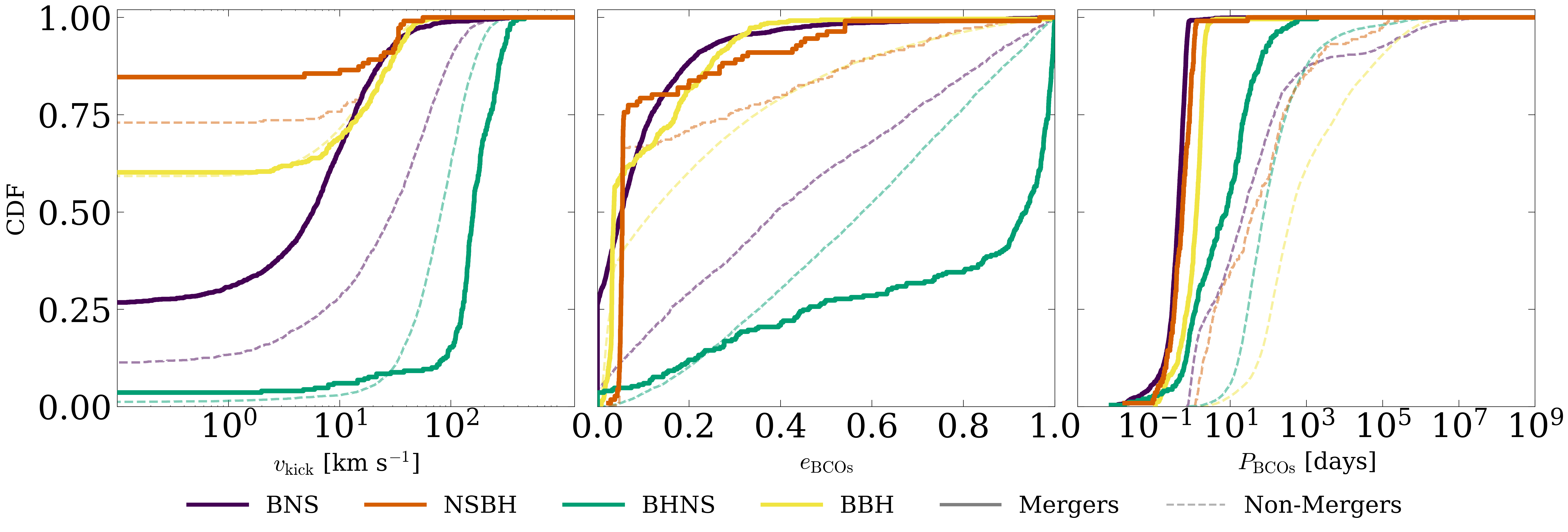}
    \caption{CDFs of key physical parameters determining the BCO merger fate: (a) second supernova kick magnitude (left panel), (b) final orbital eccentricity (middle  panel), and (c) final orbital period (right panel). The populations are divided into BNS (purple), BBH (yellow), and mixed systems (NSBH/BHNS, shown in orange/green). Solid lines represent merging systems ($t_{\mathrm{delay}} < 14$~Gyr), while dashed lines represent non-merging systems ($t_{\mathrm{delay}} > 14$~Gyr). }
    \label{fig:physical_drivers}
\end{figure*}

As expected, natal kicks impact the orbital stability of NS systems more significantly than BH systems. This is clearly observed in the leftmost panel of Figure \ref{fig:physical_drivers}, where the $v_{\mathrm{kick}}$ distributions for merging and non-merging systems are compared. For BNS, systems that successfully merge exhibit systematically lower kick magnitudes than those that fail. This result is in agreement with previous population-synthesis and observational studies showing that large NS natal kicks efficiently disrupt HMXB progenitors of BCO mergers \citep[see e.g.,][]{Igoshev2021,Wang2025}. Furthermore, while merging systems generally present lower eccentricities, this parameter is secondary to the orbital period in determining the merger timescale. The third panel shows that merging BNS are characterized by significantly shorter periods, consistent with earlier findings that orbital tightening during multiple CE phases, combined with relatively small kicks is the dominant channel leading to BNS mergers \citep[see,][]{Belczynski2002,Mapelli2018}.

In contrast with BNS, the BHNS progenitors (green lines)—originating from BH-HMXBs where the companion explodes as a NS—are subject to high natal kicks. As seen in the rightmost panel of Figure \ref{fig:physical_drivers}, these systems reside at significantly longer orbital periods ($P_{\rm BCOs} > 100$ days). However, the middle panel reveals their survival mechanism: they possess extremely high eccentricities ($e_{\rm BCOs} \gtrsim 0.9$). This confirms that for BHNS systems, a high kick that induces high eccentricity is the primary driver for merger, compensating for the wider orbital separations, these are the systems that populate the top part of the left column in Fig \ref{fig:interaction_dynamics}.

Regarding BBH systems, the role of natal kicks is largely negligible. In these systems, the kick magnitude is proportional to the ejected mass; since BHs retain most of their mass during formation (often through direct collapse where $v_{\mathrm{kick}} = 0$), the kick is effectively suppressed. This is evident in the CDF, where approximately 60\% of the BBH population exhibits zero kick velocity. 
Moreover, the nearly identical $v_{\mathrm{kick}}$ distributions for merging and non-merging BBH systems indicate that kicks are not the primary drivers of merger failure for BH. This stands in sharp contrast to the BNS population, where the second supernova kick acts as a critical filter for the final merger fate. Similar results have been reported in previous population synthesis studies when adopting fallback-modulated or vanishing BH natal kicks, finding that BBH merger efficiencies are insensitive to the natal kick prescription \citep[e.g.,][]{Fryer2012,Dominik2015,Mapelli2018}

The NSBH progenitors (orange lines) originating from an initial NS-HMXB where the massive companion collapses into a BH—follow a 'low-kick' evolutionary path. Their kick velocity distribution closely mirrors that of BBH systems, with a large fraction receiving effectively zero kicks ($v_{kick} = 0$). As a result, they are found at relatively short orbital periods and low eccentricities, driven by the direct collapse of the massive donor.

\subsubsection{The Interplay of Mass and Orbital Dynamics in Merger Efficiency}

\begin{figure}[ht]
    \centering
    \includegraphics[width=1.\columnwidth]{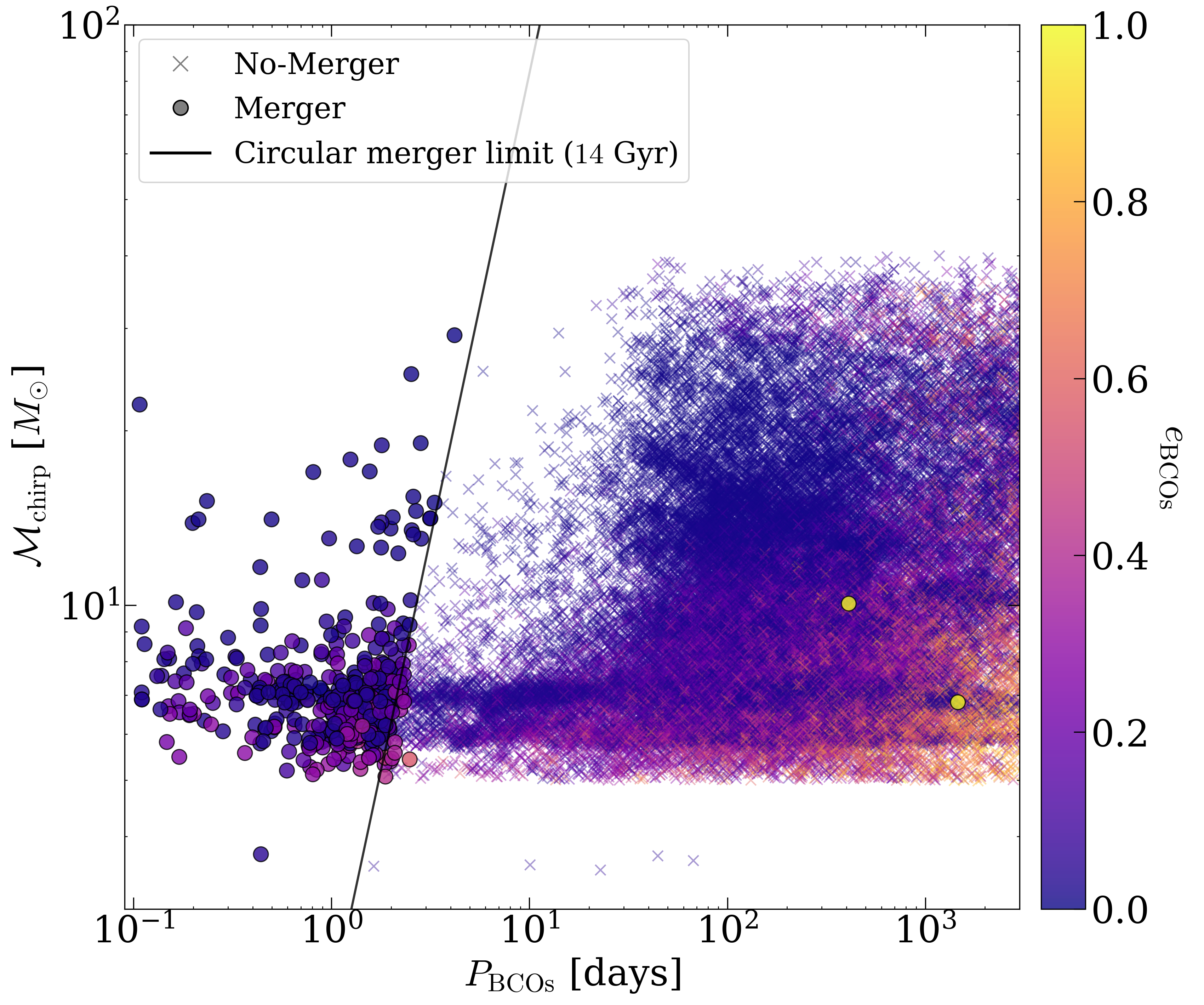}
    \caption{Distribution of the BBH population in the chirp mass ($\mathcal{M}_{\mathrm{chirp}}$) versus orbital period ($P_{\rm BCOs}$) plane at the moment of BCO formation. Circles represent merging systems ($t_{\mathrm{delay}} \leq 14$~Gyr), while crosses indicate non-merging systems ($t_{\mathrm{delay}} > 14$~Gyr). The color scale indicates the orbital eccentricity ($e_{\rm BCOs}$) at formation. The solid black line represents the theoretical limit for a circular binary to merge within a Hubble time ($14$~Gyr) according to the Peters equations  \citep{Peters1964}. Note that while merging systems are generally characterized by short orbital periods, highly eccentric systems ($e_{\rm BCOs} \gtrsim 0.9$) can merge within the age of the Universe even at periods of $P_{\rm BCOs} \sim 10^3$~days.}
    \label{fig:chirp}
\end{figure}

To identify the physical mechanisms driving the evolution of BBHs, Figure~\ref{fig:chirp} displays the population distribution in the chirp mass ($\mathcal{M}_{\mathrm{chirp}}$, as defined in Eq. \ref{eq:chirp_mass}), versus the final orbital period plane, color-coded by eccentricity.
  The data reveal a clear distinction between merging systems (circles, $t_{\mathrm{merger}} < 14$~Gyr) and non-merging systems (crosses). While both populations can achieve similarly short orbital periods ($2-3$~days) following the CE phase, their final outcomes diverge due to their mass content and specific orbital configurations. 

Since the gravitational wave inspiral time for a circular orbit is governed by the relation \citep{Peters1964}, $t_{\mathrm{merger}} \propto P^{8/3} \mathcal{M}_{\mathrm{chirp}}^{-5/3}$, this scaling implies that, for a given orbital period, more massive systems merge more rapidly. However, mass and period are not the only factors at play; the ultimate outcome arises from a combined influence of CE evolution, natal kick for newly-born BBHs, and orbital eccentricity after the formation. As illustrated in Figure ~\ref{fig:chirp}, some merging binaries occupy comparatively long orbital periods (up to $\sim 10^3$~days). These systems are pushed to merge not primarily because their orbits are tightly bound, but because supernova natal kicks excite extreme eccentricities ($e_{\rm BCOs} \sim 1$), sharply decreasing the periastron distance and accelerating gravitational wave emission.

Furthermore, the metallicity of the progenitor plays a fundamental role. High-metallicity environments are characterized by intense stellar winds that strip the progenitor stars, leaving behind lighter remnants ($\mathcal{M}_\mathrm{chirp} \lesssim 10 \, M_{\odot}$) and favoring orbital widening, which counteracts the shrinkage from the CE phase. This disadvantage restricts high-$Z$ systems in the low-mass, non-merging regime. In contrast, low-$Z$ progenitors experience reduced mass loss, allowing them to retain more massive cores ($\mathcal{M}_\mathrm{chirp} \gtrsim 20 \, M_{\odot}$) necessary to facilitate efficient orbital decay and populate the primary merger channel.

\subsection{Observational and Model Uncertainties}

The comparison between our simulated populations and the observed MW HMXBs must account for several observational limitations and biases. While X-ray surveys are essential, they are often limited by sensitivity and completeness thresholds. A major challenge arises from the heterogeneity of the observational data. Our study highlights the difficulty in performing direct comparisons when fundamental parameters—such as masses, orbital elements, and precise distances—are not available simultaneously for a single, large, and homogeneous sample. This is particularly evident in the analysis of the radial distribution (Fig.~\ref{fig:mw_properties}), where the scarcity of sources with well-constrained spatial and physical properties limits our ability to draw more definitive conclusions about the galactic distribution of these systems. Furthermore, the intrinsic variability of HMXBs, introduces a temporal selection effect. This variability complicates the modeling of a representative X-ray luminosity, potentially skewing the observed orbital and mass distributions compared to the underlying physical population \citep[see e.g.,][]{Reig2013,Binder2025}.

Finally, regarding stellar evolution models, significant uncertainties remain in the treatment of binary interactions. The physics of CE evolution and the efficiency of SN engines are among the most critical unknowns. Furthermore, mass-transfer efficiency and angular momentum loss may be also important to consider. In this work, we adopted specific analytical prescriptions and, in some cases, implemented hard-coded distributions for certain parameters (as detailed in Section~\ref{ref:IC}). An additional source of uncertainty concerns the observability of wind-accreting BH-HMXBs. \citet{Hirai2021} show that the formation of an accretion disk-required for significant X-ray emission, depends critically on the Roche-lobe filling factor of the donor star, $f$, with disk formation expected only when $f\gtrsim 0.8-0.9$. Since our current model does not impose this condition, a fraction of the wind-fed BH-HMXBs in our synthetic population may not produce detectable X-ray emission, which could partially account for the known discrepancy between large theoretical BH-HMXB populations and the small number of observed systems. While these assumptions provide a necessary framework for this study, a comprehensive exploration of the parameter space of these population models—including the effects of binary evolution uncertainties, accretion disk formation criteria, and the extension from wind-fed systems to RLOF HMXBs and Be X-ray binaries—will be discussed in a forthcoming paper (Vivanco et al., in prep.). 

\section{Summary and Conclusions}\label{sec:conclusions}

In this study, we investigate the wind-fed HMXB population of the MW by integrating the population synthesis code {\sc sevn} with a set of 66 MW-like galaxies from the TNG50 cosmological simulation.
Our method offers a realistic scheme for modelling HMXBs in the MW by explicitly incorporating the diverse metallicities and spatial distributions of binary systems. Our main findings are summarized as follows:

\begin{itemize}
    
    \item The simulated HMXB population closely traces the galactic spiral arms, a direct result of their link to young stellar particles ($< 40$ Myr, see Fig. \ref{fig:mw_properties}). Furthermore, BH-HMXBs tend to be younger systems ($< 10$ Myr) originating from binaries with a primary mass $M_{ZAMS,0} \gtrsim 20 M_{\odot}$, while NS-HMXBs are typically older ($> 10$ Myr),  associated with a lower-mass primary ($\sim 8-25 M_{\odot}$).

    \item We find that about $\sim 0.2-3.2\%$ of BH-HMXBs eventually form BCO mergers (BBH and BHNS) within a Hubble time, whereas for NS-HMXBs this fraction is $\sim 3.6-23.4\%$. If we restrict the population by X-ray luminosities in the range $10^{35} \le L_X \le 10^{40}$ erg s$^{-1}$, we obtain that about $\sim 0.3-5.4\%$ of BH-HMXBs produce BBH or BHNS mergers, and $\sim 3.5-26.0\%$ of NS-HMXBs evolve into BCO mergers, including BNS and NSBH systems. 

    \item We identified distinct orbital regions for future GW events. Merger candidates from the BH-HMXB population are primarily confined to orbital periods $P_{\rm HMXBs} \le 10^3$ days and low to moderate eccentricities ($e_{\rm HMXBs} \le 0.4$). On the other hand, the NS-HMXB merger progenitor region extends to  periods ($P_{\rm HMXBs} \le 10^4$ days) but encompasses a significantly broader range of eccentricities, reaching up to $e_{\rm HMXBs} \approx 0.9$ (Fig. \ref{fig:period_eccentricity_hmxbs}).
    
    \item Pre-HMXB evolution acts as a first filter for future GW events. Systems merging within a Hubble time ($t_{\mathrm{delay}} < 14$ Gyr) are dominated by those that underwent significant orbital hardening through CE phases (CIII and IV). While SMT ensures binary survival, it often leads to wide orbits that prevent future coalescence unless high eccentricities are induced by natal kicks (Fig. \ref{fig:dominant_channels_p_e} and \ref{fig:interaction_dynamics}).
    
    \item For the BNS population, the magnitude of the second supernova natal kick serves as a filter, with merging systems exhibiting systematically lower kicks and shorter periods (Fig. \ref{fig:physical_drivers}). In contrast, mixed BHNS systems occupy a distinct region characterized by long orbital periods ($P_{\rm BCOs} > 100$ days) and high natal kicks; their merger is primarily enabled by extreme eccentricities ($e_{\rm BCOs} \gtrsim 0.9$) that significantly reduce the coalescence time. Finally, for BBH systems, the merger fate is primarily driven by the interplay between chirp mass and orbital period; low-metallicity progenitors are more likely to populate the primary merger channel by preserving the heavy cores necessary for efficient gravitational wave-driven inspiral.
\end{itemize}

By populating star-forming particles with HMXBs using \textsc{sevn} and TNG50, we obtain detailed spatial and demographic predictions for the Galactic HMXB population. We can naturally recover their distribution along galactic spiral arms, and age distributions \citep{Fortin2022}. We further identify the orbital configurations and evolutionary pathways that enable an HMXB to evolve into a BCO merger within a Hubble time. Our results indicate that only a small fraction of HMXBs reach this stage: while the HMXB phase can precede BCO formation, it does not imply that it leads to a GW event. In particular, substantial orbital hardening through binary interactions is necessary but not sufficient for merger, as many systems that survive mass transfer or CE evolution still end up in orbits too wide to merge. Future work exploring the uncertain physics of CE evolution, mass transfer and natal kicks will be crucial to tighten these constraints and connect the observed HMXB population to the GW merger population in a fully statistical way.

\begin{acknowledgements}
FV and MCA acknowledge financial support from Fondecyt Iniciación number 11240540 and ANID BASAL project FB210003. BL gratefully acknowledges the funding of the Deutsche Forschungsgemeinschaft (DFG, German Research Foundation) under Germany's Excellence Strategy EXC 2181/1 - 390900948 (the Heidelberg STRUCTURES Excellence Cluster). GI was supported by a fellowship grant from the la Caixa Foundation (ID 100010434). The fellowship code is LCF/BQ/PI24/12040020. This research was partially supported by the supercomputing infrastructure of the NLHPC (CCSS210001). This work made partial use of the RAGNAR computer server at Universidad Andrés Bello for calculations. This work was funded by the National Agency for Research and Development (ANID) / Scholarship Program / DOCTORADO NACIONAL / 2021 - 21222248. FV acknowledges the Department of Physics and Astronomy at the University of Padova for its hospitality, where part of this work was carried out during a research stay supported by ANID.
\end{acknowledgements}
%
\bibliographystyle{aa} 
\bibliography{references} 
\appendix
\section{Efficiency}\label{ap:efficiencies}
\begin{figure}[h!]
    \centering
    \includegraphics[width=\linewidth]{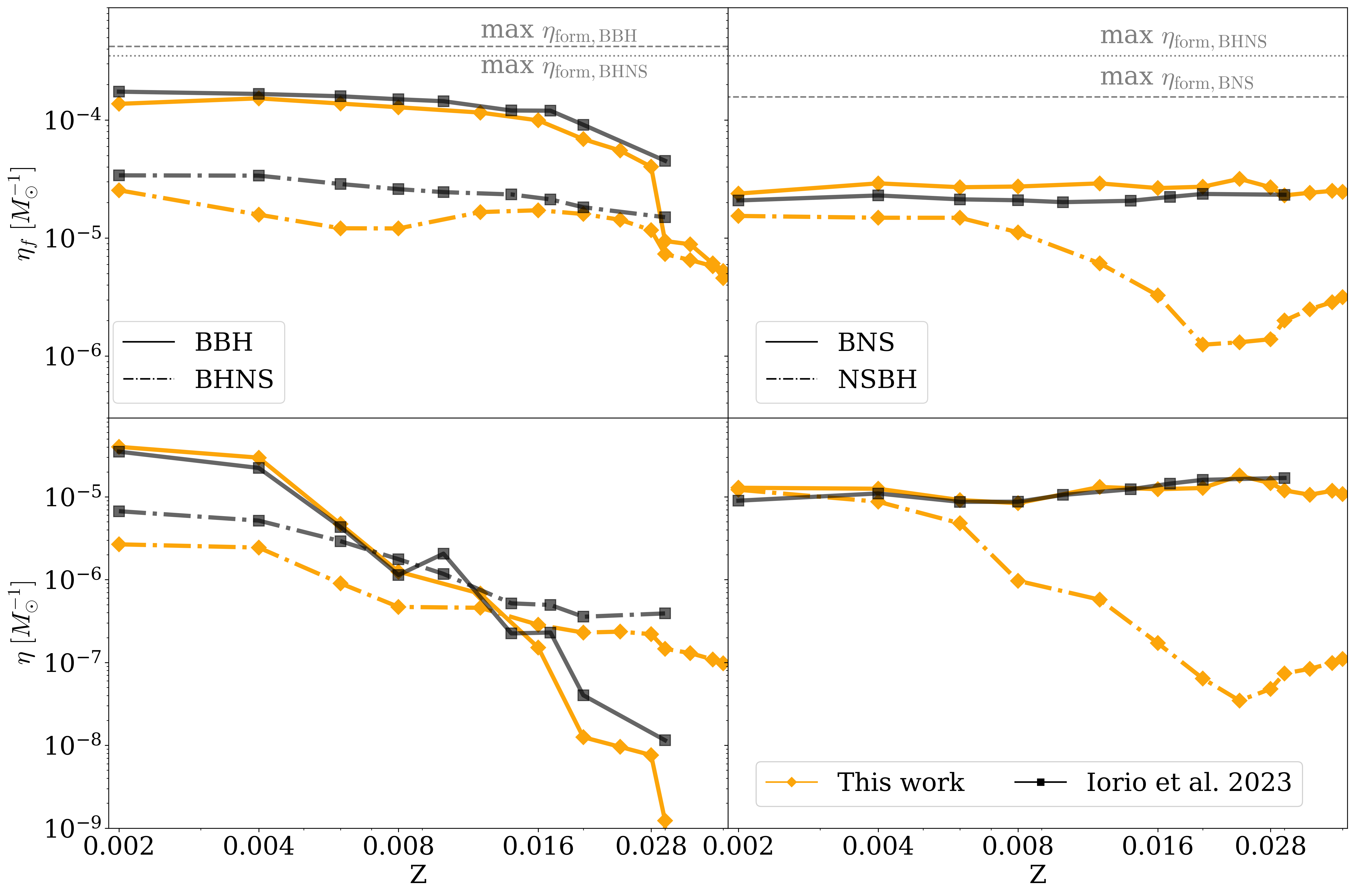} 
    \caption{Formation efficiency ($\eta_f$, top panels) and merger efficiency ($\eta$, bottom panels) as a function of metallicity ($Z$) for BCOs. The left column shows systems originating from BH-HMXB progenitors, whereas the right column displays those from NS-HMXB progenitors. Orange lines represent our updated single-model results (incorporating the \texttt{Rapid-DG} supernova prescription and conservative mass transfer, $f_{\mathrm{MT}}=1.0$), strictly tracking binaries that pass through an observable HMXB phase. Black lines denote the baseline population synthesis results from \citet{Iorio2023}. Solid lines indicate BBH (left) or BNS (right) systems, while dash-dotted lines correspond to mixed systems. Note that our model explicitly separates BHNS and NSBH evolutionary paths, whereas the baseline model combines them. Grey dashed and dotted lines mark the theoretical maximum formation efficiencies derived by \citet{vanSon2025}.}
    \label{fig:HMXB_efficiency}
\end{figure}

Figure~\ref{fig:HMXB_efficiency} displays the formation efficiency ($\eta_f$, top panels) and the merger efficiency ($\eta$, bottom panels) for BCOs from BH-HMXB and NS-HMXB progenitors, defined as:

\begin{equation}
    \eta_f = \frac{N_{\mathrm{BCO}}\,}{M_{\mathrm{pop}}},
    \qquad
    \eta = \frac{N_{\mathrm{BCO}}(t_{\mathrm{delay}} < 14~\mathrm{Gyr})\,\, }{M_{\mathrm{pop}}},
\end{equation}

where $N_{\mathrm{BCO}}$ is the total number of BCOs formed and $N_{\mathrm{BCO}}(t_{\mathrm{delay}} < 14~\mathrm{Gyr})$ is the subset merging within a Hubble time. The term $M_{\mathrm{pop}}$ represents the effective total mass of $M_{\mathrm{pop}} = 4.07 \times 10^{9} \, \mathrm{M}_\odot$, assuming a binary fraction $f_{\mathrm{bin}} = 0.5$ \citep{Sana2012} and applying a correction factor (\(f_{\mathrm{corr}} = 0.172\)) for the incomplete IMF sampling due to our mass limits.

In our analysis, we compare our single model configuration (orange lines) against the fiducial model from \citet{Iorio2023} (black lines). The main variations between our configurations are that \citet{Iorio2023} adopt a mass transfer efficiency of $f_{\mathrm{MT}}=0.5$ and the standard \texttt{rapid\_gauNS} supernova model, while our setup assumes conservative mass transfer ($f_{\mathrm{MT}}=1.0$). Additionally, we implement the supernova prescription (\texttt{Rapid-DG}) as a first approximation to better reproduce the Galactic NS mass distributions observed by the LVK collaboration \citep{Abbott2023}, thereby yielding more massive NSs. 

For binary black holes (BBH, solid lines in the left panels), our formation efficiency is slightly lower than that of \citet{Iorio2023}. This difference is expected, as our sample strictly isolates systems that successfully transition through an HMXB phase in the full range ($10^{31}-10^{40}$ erg/s), naturally filtering the broader BBH population. Both models remain well below the theoretical maximum formation efficiency derived by \citet{vanSon2025}. We observe a sharp decline in BBH efficiency beyond $Z \approx 0.016$ as a result of line-driven winds that reduce core masses, leading to larger natal kicks or triggering premature stellar mergers.

For BNS (solid lines in the right panels), the formation efficiency remains largely insensitive to metallicity. This is because NS progenitors are lower-mass stars and are less affected by $Z$-dependent winds. Our BNS formation efficiency (top right panel) is slightly higher than that of \citet{Iorio2023}, which can be attributed to our updated model parameters.

A notable difference between our work and \citet{Iorio2023} is the treatment of mixed systems. While they combined BHNS and NSBH into a single category, we separate them based on their formation history (dashed-dotted lines). Consequently, our NSBH formation efficiency (top right) shows a decline at higher metallicities, a direct result of the severe mass loss experienced by the BH progenitor in the system. 

Regarding merger efficiency (bottom panels), BBH systems show a steeper decline than formation efficiency at high $Z$. Severe mass loss results in lower remnant masses, which significantly increases the gravitational wave merger timescale. This timescale is dependent on the chirp mass, defined as:

\begin{equation}\label{eq:chirp_mass}
    \mathcal{M}_{\mathrm{chirp}} = \frac{(M_1 M_2)^{3/5}}{(M_1 + M_2)^{1/5}}
\end{equation}

Since the merger timescale follows $t_{\mathrm{merger}} \propto P^{8/3} \mathcal{M}_{\mathrm{chirp}}^{-5/3}$, the combination of lower remnant masses and wider orbits delays coalescence considerably. For NSNS systems, however, $\eta \approx \eta_f$ across all metallicities, implying that systems surviving the second supernova kick almost always remain in orbits tight enough to merge within a Hubble time.

\section{Properties of HMXBs according to age}\label{appendix:b2}

\begin{figure}[h!]
    \centering
    \includegraphics[width=\linewidth]{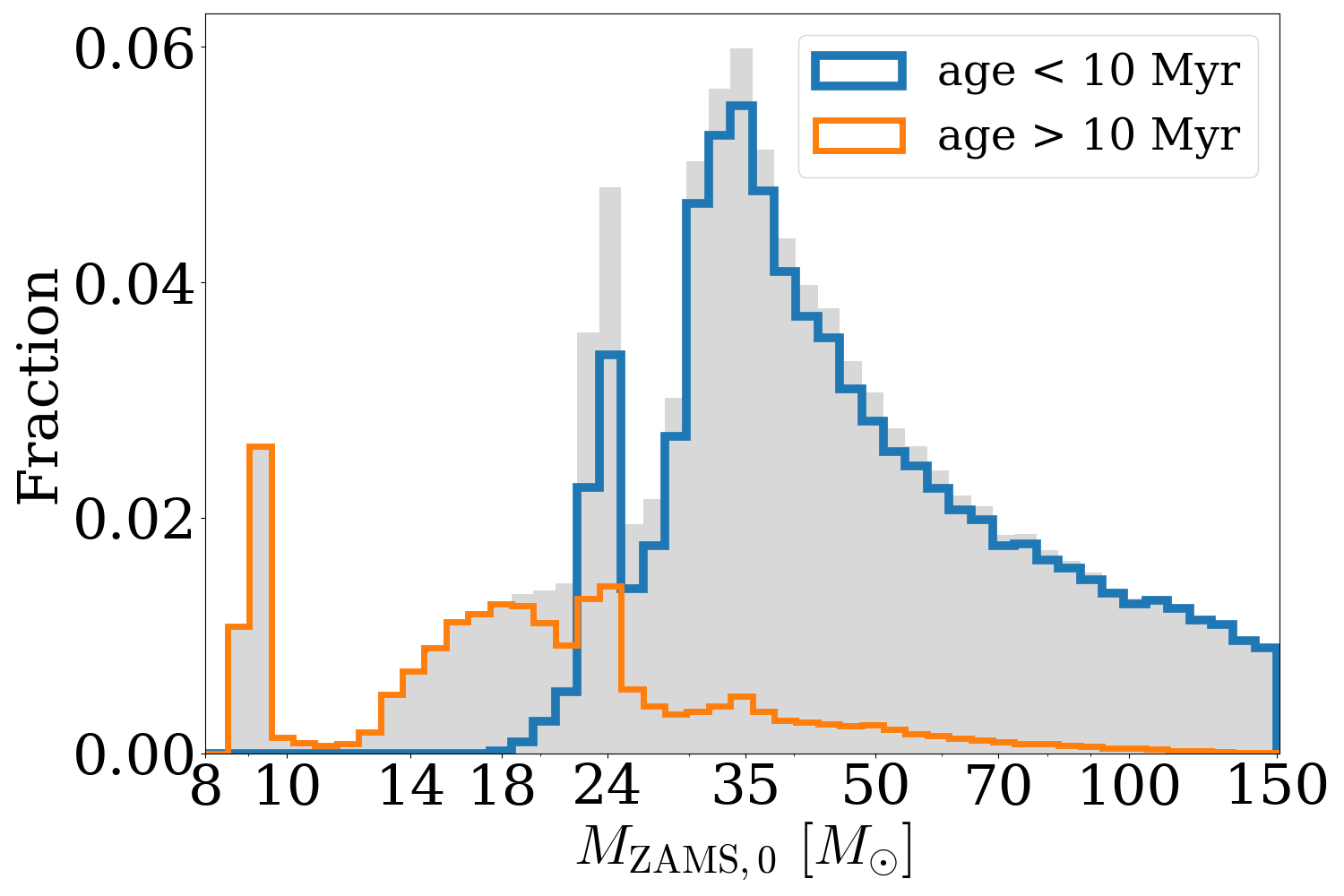} 
    \caption{Normalized distribution of the initial ZAMS mass of the primary star ($M_{\mathrm{ZAMS},0}$) for the synthetic HMXB population accumulated across all sampled galaxies. The filled gray area represents the total population. The sub-populations are divided by the system's age: younger systems are shown with the blue line, while older systems are shown with the orange line.}
    \label{fig:mzams_distribution}
\end{figure}

\begin{figure}[h!]
    \centering
    \includegraphics[width=\linewidth]{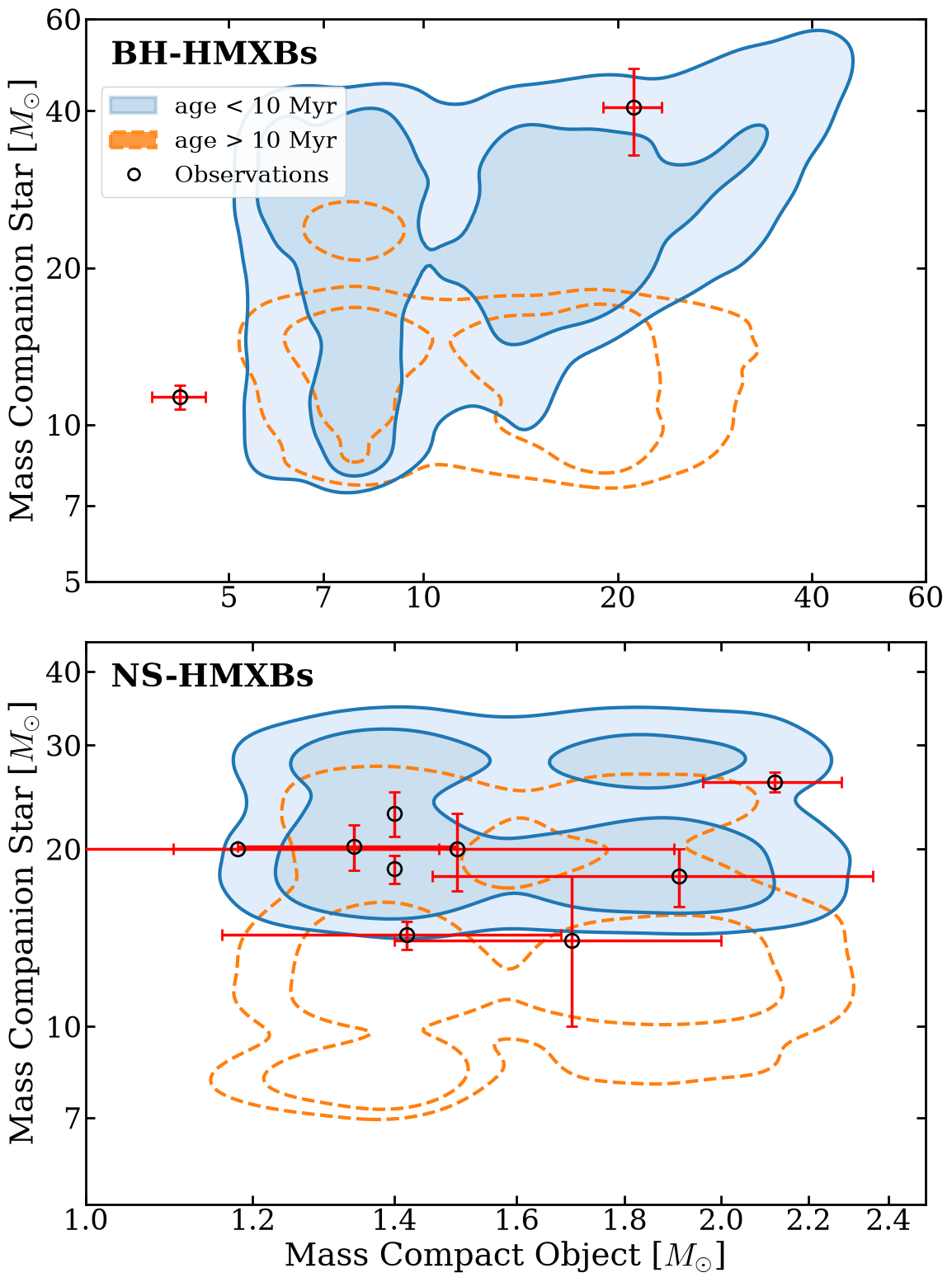}
    \caption{Component mass distributions for the synthetic HMXB populations, estimated via KDE. The top panel corresponds to BH-HMXBs, displaying the companion star mass versus the compact object mass. The bottom panel shows the equivalent distribution for NS-HMXBs. Contours indicate the $1\sigma$ and $2\sigma$ (68\% and 95\%) confidence intervals. The filled blue contours represent the density of younger systems ($\text{age} < 10 \text{ Myr}$), while the dashed orange contours trace the older population ($\text{age} > 10 \text{ Myr}$). Red circles with error bars denote observational data from Fortin et al. (2023)}
    \label{fig:mass_distribution_kde}
\end{figure}

In this section, we present a detailed analysis of the HMXB population properties by segregating the systems based on their evolutionary age. By examining the initial masses of the primary stars ($M_{\mathrm{ZAMS},0}$), we find a clear mass segregation (see Fig. \ref{fig:mzams_distribution}): the peak observed at ages $< 10$~Myr originates primarily from stars with initial masses $\gtrsim 20 M_{\odot}$, whereas the population at $> 10$~Myr is dominated by progenitors with masses between 8 and $25 M_{\odot}$. This segregation directly determines the type of compact object hosted; as shown in the component mass distributions in Fig. \ref{fig:mass_distribution_kde}, the majority of the BH-HMXB population is found at ages $< 10$~Myr. Only a small fraction of BH-HMXBs exceeds 10~Myr, and these older systems are characterized by companion stars with masses $< 20 M_{\odot}$. Conversely, the majority of the NS-HMXB population is found in the $> 10$~Myr range, associated with companion stars $\lesssim 25 M_{\odot}$ and a higher concentration between 8 and $15 M_{\odot}$.

Regarding the comparison with observational data, we overlay the Galactic HMXB sample from \citet{Fortin2023} in Fig. \ref{fig:mass_distribution_kde}. We note that the current observational sample is relatively small, which makes it challenging to discern a definitive trend between system age and component masses. For the NS-HMXB population (bottom panel of Fig. \ref{fig:mass_distribution_kde}), the observed systems are predominantly located in the region where the younger (solid blue) and older (dashed orange) KDE contours overlap. This spatial coincidence, combined with the limited number of sources, prevents a clear statistical distinction of age sub-populations based solely on their position in the mass-companion star mass plane. However, the observed Galactic systems are consistent with the high-density regions predicted by our synthetic models. Kinematic (post-SN) age estimates for several of these systems, as reported in \citet{Fortin2022}, align with the post-supernova temporal ranges produced in our simulations (as discussed in Sect. \ref{HMXBs_population}), although they are subject to large individual astrometric uncertainties.

\begin{figure}[h!]
    \centering
    \includegraphics[width=\linewidth]{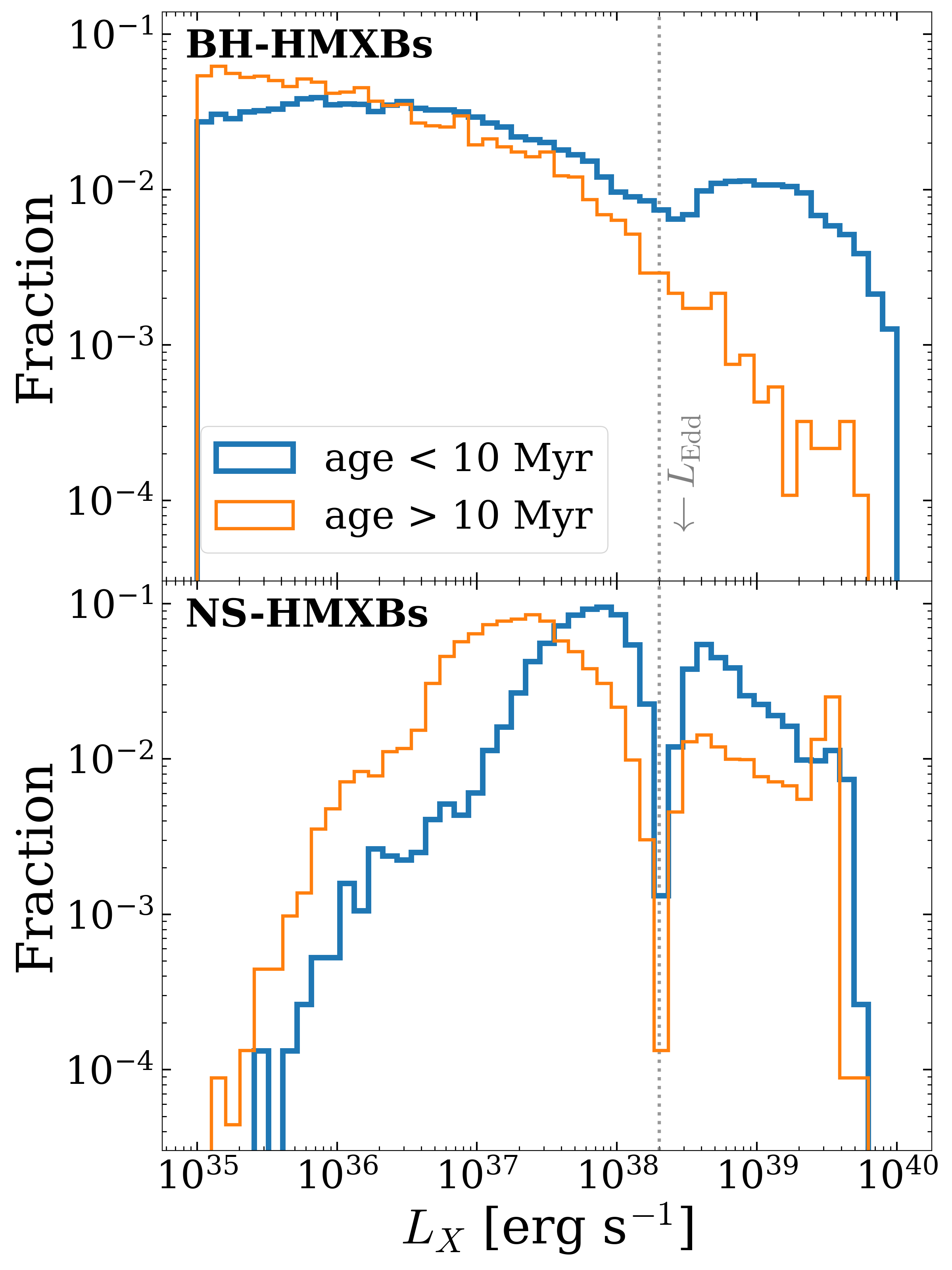}   
    \caption{Normalized X-ray luminosity distributions ($L_X$) for the synthetic HMXB population. The top panel displays BH-HMXBs, while the bottom panel shows NS-HMXBs. The population is divided into younger systems (blue lines) and older systems (orange lines). The vertical dotted gray line marks the Eddington luminosity for a NS ($L_{\rm Edd} \approx 2 \times 10^{38}$\,erg\,s$^{-1}$).}      
    \label{fig:app_lx_distribution}
\end{figure}

X-ray luminosity trends also vary significantly with age, as illustrated in Fig. \ref{fig:app_lx_distribution}. For BH-HMXBs, younger systems span a broad range ($10^{35} \text{--} 10^{40} \text{ erg s}^{-1}$) peaking at $\sim 10^{36} \text{ erg s}^{-1}$, while older systems are generally fainter. In the case of NS-HMXBs, the younger population peaks at a higher luminosity ($\sim 10^{38} \text{ erg s}^{-1}$) than the older one ($\sim 2 \times 10^{37} \text{ erg s}^{-1}$). Notably, both populations show a high-luminosity tail extending beyond the standard Eddington limit ($L_{\text{Edd}}$), which is a consequence of the super-Eddington accretion factor of 20 implemented in the SEVN code.

\section{Formation Channels for HMXBs}\label{appendix:channels}
Following the classification by \citet{Iorio2023} \citep[see also,][]{Broekgaarden2021}, in Section~\ref{pre_HMXBs}, we categorize the evolutionary pathways leading to merging BCO into four main channels and two non-interacting scenarios. The details of each channel are described as follows:

\begin{itemize}
    \item Channel I (SMT + CE, CI): Systems undergoing SMT before the first compact object forms, followed by at least one CE phase. This is the standard formation channel.
    \item Channel II (SMT only, CII): Systems evolving exclusively through SMT, avoiding any CE phase.
    \item Channel III (Early CE, Single Core, CIII): Systems experiencing a CE phase \textit{before} the first remnant forms. At the time of formation, the binary consists of one H-rich star and one stripped (pure-He or naked-CO star) star.
    \item Channel IV (Early CE, Double Core, CIV): Similar to Channel III, but both stars lose their hydrogen envelopes via CE prior to the first remnant formation.
    \item Channels 0 and V (Non-interacting, C0 and CV): Rare cases where the binary avoids mass transfer either entirely (C0) or prior to the first remnant formation (CV), relying on favorable natal kicks to harden the orbit for a subsequent merger.
\end{itemize}
\end{document}